\documentclass{aastex63}

\shorttitle{Asymmetric Disk of DM Tau}
\shortauthors{Hashimoto et al.}

\graphicspath{{./}{figures/}}

\begin{document}

\title{ALMA Observations of the Asymmetric Dust Disk around DM Tau}

\correspondingauthor{Jun Hashimoto}
\email{jun.hashimto@nao.ac.jp}

\author[0000-0002-3053-3575]{Jun Hashimoto}
\affil{Astrobiology Center, National Institutes of Natural Sciences, 2-21-1 Osawa, Mitaka, Tokyo 181-8588, Japan}
\affil{Subaru Telescope, National Astronomical Observatory of Japan, Mitaka, Tokyo 181-8588, Japan}
\affil{Department of Astronomy, School of Science, Graduate University for Advanced Studies (SOKENDAI), Mitaka, Tokyo 181-8588, Japan}

\author{Takayuki Muto}
\affil{Division of Liberal Arts, Kogakuin University, 1-24-2, Nishi-Shinjuku, Shinjuku-ku, Tokyo, 163-8677, Japan}

\author[0000-0001-9290-7846]{Ruobing Dong}
\affil{Department of Physics \& Astronomy, University of Victoria, Victoria, BC, V8P 1A1, Canada}

\author[0000-0003-2300-2626]{Hauyu Baobab Liu}
\affiliation{Institute of Astronomy and Astrophysics, Academia Sinica, 11F of Astronomy-Mathematics Building, AS/NTU No.1, Sec. 4,
Roosevelt Rd, Taipei 10617, Taiwan, ROC}

\author[0000-0003-2458-9756]{Nienke van der Marel}
\affil{Department of Physics \& Astronomy, University of Victoria, Victoria, BC, V8P 1A1, Canada}

\author[0000-0001-8822-6327]{Logan Francis}
\affil{Department of Physics \& Astronomy, University of Victoria, Victoria, BC, V8P 1A1, Canada}

\author{Yasuhiro Hasegawa}
\affil{Jet Propulsion Laboratory, California Institute of Technology, Pasadena, CA 91109, USA}

\author[0000-0002-6034-2892]{Takashi Tsukagoshi}
\affil{Division of Science, National Astronomical Observatory of Japan, Mitaka, Tokyo 181-8588, Japan}

\begin{abstract}

We report an analysis of the dust disk around DM~Tau, newly observed with the Atacama Large Millimeter/submillimeter Array (ALMA) at 1.3 mm. The ALMA observations with high sensitivity (8.4~$\mu$Jy/beam) and high angular resolution (35~mas, 5.1~au) detect two asymmetries on the ring at $r\sim$20~au. They could be two vortices in early evolution, the destruction of a large scale vortex, or double continuum emission peaks with different dust sizes. We also found millimeter emissions with $\sim$50~$\mu$Jy (a lower limit dust mass of 0.3~$M_{\rm Moon}$) inside the 3-au ring. To characterize these emissions, we modeled the spectral energy distribution (SED) of DM~Tau using a Monte Carlo radiative transfer code. We found that an additional ring at $r=$ 1~au could explain both the DM~Tau SED and the central point source. The disk midplane temperature at the 1-au ring calculated in our modeling is less than the typical water sublimation temperature of 150~K, prompting the possibility of forming small icy planets there.
\end{abstract}


\section{Introduction} \label{sec:intro}

Planets are believed to form in protoplanetary disks \citep[e.g.,][]{haya85,poll1996}. The early stages of planet formation can be identified by disk structures such as gaps and asymmetric structures via planet--disk interactions \citep[e.g.,][]{kley12}. Such structures have been reported in tens of dust disks with Atacama Large (sub-)Millimeter Array \citep[ALMA; e.g.,][]{vandermarel18,andr2018,feng2018a}, and roughly 10 disks show asymmetries \citep[e.g.,][]{fran2020a,vandermarel20a,tsukagoshi2019a,perez2018a,dong2018mwc758}. Particularly, asymmetric structures such as the blob and the crescent features possibly due to azimuthal gas pressure maxima \citep[e.g.,][]{raettig2015,ragusa2017} could be on-going planet forming sites because gas pressure maxima efficiently trap dust grains, potentially leading to planetesimal formation. Two major possible origins of these asymmetries  are discussed by \citet{vandermarel20a}: long-lived anticyclonic vortices at gap edges possibly curved by companions \citep[e.g.,][]{raettig2015} and gas horseshoes due to eccentric cavities curved by massive companions \citep[e.g.,][]{ragusa2017}. The main difference between the two is in the mass of companions: a vortex can be produced at edges of gaps possibly opened by planets, whereas a horseshoe structure needs to be triggered by a much more massive companion, i.e., a brown dwarf. Therefore, in the vortex scenario, asymmetries could be signpost of planets, while they are not in the horseshoe scenario. Though the origins of individual asymmetric disks have not been determined by current observations \citep{vandermarel20a}, investigating asymmetric disks could help understanding of planet formation.

DM~Tau (spectral type: M1, \citealp{keny95}; $T_{\rm eff}$: 3705~K, \citealp{andr11}; $M_{\rm *}$: 0.53~$M_{\odot}$, \citealp{piet07}; distance: 145~pc, \citealp{gaia18}) is a single star system \citep{nguyen2012,will16}, and its protoplanetary disk has a weak asymmetry in the outer disk at $r\sim$20~au \citep{kudo2018}. The spectral energy distribution (SED) of DM~Tau shows a deficit at $\lambda\sim$1--10~$\mu$m, which was interpreted as the presence of a deep cavity at $r\sim$ 3~au around DM~Tau by analyzing its SED \citep[e.g.,][]{calvet+02}. Subsequent sub-millimeter interferometric observations with a beam size of 0$\farcs$3 detected a 20-au dust ring \citep{andr11}. These discrepancies were explained by recent ALMA long-baseline observations \citep{kudo2018}: DM~Tau has multiple rings at $r=$3 and 20~au and low contrast rings at $r\gtrsim$ 60~au. The $^{12}$CO~(2--1) gas disk around DM~Tau has no cavity/ring structures \citep{kudo2018}, possibly due to a high optical depth of $^{12}$CO, while other molecular species such as C$_{2}$H show a ring structure at $r\sim$ 80~au \citep{berg16}. A candidate giant planet was reported at $r\sim$ 6~au by near-infrared sparse aperture masking interferometry \citep{will16}, but this detection has not yet been confirmed. The mass accretion rate of DM~Tau, $\dot{M}$~$\sim$ 6~$\times $ 10$^{-9}$~$M_{\odot}$/yr \citep{mana14}, is comparable with that of typical T~Tauri~stars \citep{naji15}. As small (sub-micron size) dust grains coupled with the gas flow into the central star, significant infrared excess at $\lambda\sim$1--10~$\mu$m should appear in the SED. Hence the origin of DM~Tau's 3-au cavity with both high mass accretion rate and strongly depleted dust grains in the cavity is still under debate \citep[e.g.,][]{mana14,kudo2018}. 

In this paper, we report follow-up observations of DM~Tau with ALMA in cycle~6. Our original aim for these new observations was to confirm a weak asymmetry with a contrast of 20~\% in the inner ring at $r\sim$3~au reported by \citet{kudo2018}. Though we have not confirmed this asymmetry with our new cycle~6 observations, two blobs at the same radial location were newly identified in the outer ring at $r\sim$20~au. We also performed SED fitting to test the existence of dust grains inside the 3-au cavity around DM~Tau.

\section{Observations} \label{sec:obs}

The ALMA observations of DM~Tau summarized in Table~\ref{tab:table3} were carried out with Band 6 in the C43-9/10 configuration on 2019 June 5, UT under the project 2018.1.01755.S, using 44 antennas with a baseline length extending from 83.1~m to 15.2~km. Since the short baseline data are available in the ALMA archive (ID: 2013.1.00498.S; PI: L.~Perez), we did not request these observations. The long baseline data were taken with four spectral windows (SPWs): three with 128 channels spanning 1.875~GHz (31.25~MHz per channel) centered on 213.5, 216.3, and 228.0~GHz; and one with 3840 channels spanning 1.875~GHz (448.3~kHz per channel, 0.64~km/s velocity resolution) centered on the $^{12}$CO~$J=2\rightarrow1$ rest frequency of 230.538~GHz. The bandpass and flux calibrator were the quasar J0423$-$0120, while the phase calibrator was J0431$+$1731. The mean precipitable water vapor was 0.8--0.9~mm during observations. The total on-source integration time was 134.3 minutes. The data were calibrated by the Common Astronomy Software Applications (CASA) package \citep{mcmu07} version 5.4.0-70, following the calibration scripts provided by ALMA. We separately conducted self-calibration of the visibilities in long and short baseline data. For long baseline data, the phases were iteratively self-calibrated on solution intervals of 360~s combining all SPWs. However, as the self-calibration degraded the signal-to-noise ratio due to the large amount of flagged data, we decided not to use self-calibrated long baseline data in this paper. For short baseline data, the phases were self-calibrated once with fairly long solution intervals (solint=`inf') that combined all spectral windows.

We combined our long baseline data with short baseline data to recover the missing emission at larger angular scales. We compared the visibility amplitudes at less than 200~k$\lambda$ between the two data sets and confirmed their consistency. The images of both data sets were aligned by two manners as follows\footnote{We originally attempted to align short and long baseline data by correcting the proper motion. The proper motions of both data were calculated with the function EPOCH PROP in GAIA ADQL ({\sf https://gea.esac.esa.int/archive/}). The phase centers and pointing tables for both data sets were corrected by fixvis and fixplanets, respectively, in the CASA tools. However, we found that the new phase centers of short and long baseline data are shifted to $\sim$16 and $\sim$5~mas relative to the centers of ellipse isophoto fitting. Hence, we decided not to use the phase centers derived by correcting the proper motion.}. 

\emph{Method~A ---} We separately synthesized the dust continuum images of short and long baseline data by CASA with the \verb#CLEAN# task using a multi-frequency deconvolution algorithm \citep{rau11}. We then conducted ellipse isophoto fitting at 30~$\sigma$ in the images of both data. The phase centers for both data sets were corrected to the centers of ellipse isophoto fitting by \verb#fixvis# in the CASA tools. To test whether the new phase center is the center of the disk, we subtracted the 180\degr-rotated image in the visibility domain. This procedure corresponds to producing a synthesized image with only the imaginary part of the visibilities. Because the visibility is complex conjugate, the subtraction of the 180\degr-rotated image is mathematically equal to setting the real part as zero and doubling the value of the imaginary part, respectively. In other words, the real part contains information of both symmetric and asymmetric structures of objects, whereas the imaginary part contains only information of asymmetries. Therefore, by synthesizing the image with only the imaginary part, we selectively remove only symmetric structures, vice versa, only asymmetric structures can be efficiently detected\footnote{This method would also serve as a diagnosis to test a misalignment between an observed disk and a modeled disk. When the modeled disk is misaligned to the observed disk, spurious asymmetries could easily generate even if both disks are symmetric.}. We searched the minimum r.m.s in the central region of the images with shifting images relative to the center of ellipse isophoto fitting in the visibility domain by the phase shift defined as $\exp\left[2\pi i \left(u\,\Delta {\rm R.A.} +v\,\Delta {\rm DEC}\right)\right]$, where $u$ and $v$ are the spatial frequencies and $\Delta$R.A. and $\Delta$DEC are shift values, respectively. Figures~\ref{figA9} and \ref{figA10} in Appendix shows dust continuum images synthesized with only the imaginary part, including the image with the minimum r.m.s. We found that the shift values with the minimum r.m.s are ($\Delta$RA, $\Delta$DEC) of (0~mas, 0~mas) and ($+$4~mas, $-$2~mas) relative to the center of ellipse isophoto fitting in the long and the short baseline data, respectively. Finally, the pointing tables for both data sets were corrected toward the images with the minimum r.m.s by \verb#fixplanets# in the CASA tools. The new phase centers\footnote{The original phase centers in long and short baseline data are (04h33m48.734901s, 18d10m09.63258s) in ICRS and (04h33m48.729253s, 18d10m09.78982s) in FK5~J2000.0, respectively.} of long and short baseline data in ICRS coordinates are (4h33m48.74961s, 18d10m9.6177s) and (4h33m48.74792s, 18d10m9.6819s), respectively.

\emph{Method~B ---} We also check the shift value with the minimum $\chi^{2}$ of the imaginary part. The value of $\chi^{2}$ is defined as $\sum\,\left(W_{j}\,{\rm Im}V_{j}^{2}\right)$, where the subscript $j$ represents the $j$-th data. ${\rm Im}V_{j}$ and $W_{j}$ are the visibilities in the imaginary part and weights, respectively. We found that the shift values with the minimum $\chi^{2}$ are ($\Delta$RA, $\Delta$DEC) of ($+$1~mas, $+$1~mas) and ($+$5~mas, $-$2~mas) relative to the center of ellipse isophoto fitting in the long and the short baseline data, respectively. The method~A measures r.m.s in the region where we are interested in, while the method~B measures r.m.s of entire visibilities in the imaginary part, and thus, we rely on the results of method~A in this paper.

\begin{figure}[ht!]
\begin{centering}
\includegraphics[clip,width=\linewidth]{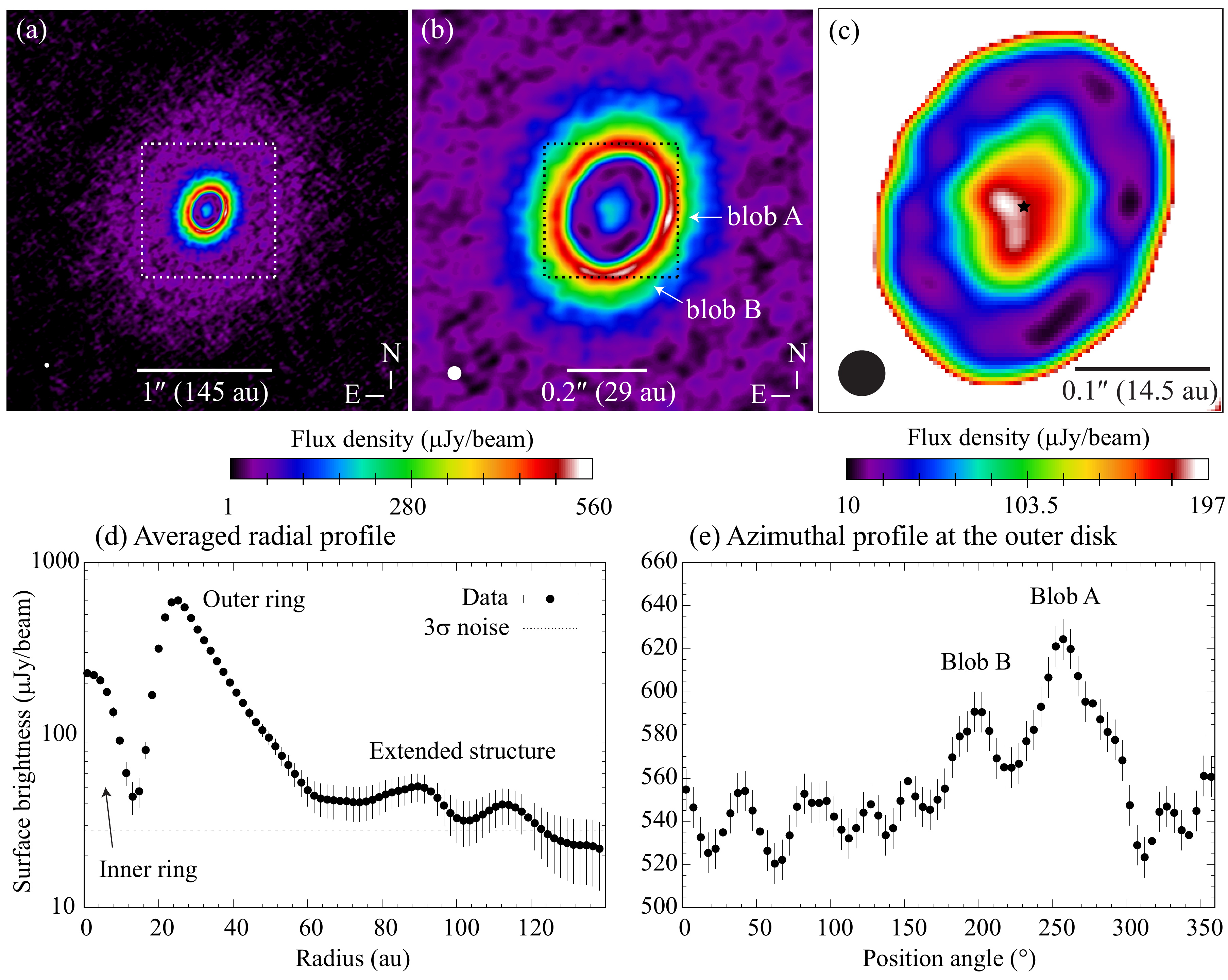}
\end{centering}
\caption{Synthesized images of the dust continuum of the DM~Tau disk obtained with ALMA cycle~6 at band~6. (a): Entire image. (b): Central magnified image of panel~(a). (c): Central magnified image of panel~(b). The regions of panel~(b) and (c) are indicated by the white and black dotted square in panel~(a) and (b), respectively. The r.m.s.\ noise measured in the region far from the object is 8.4~$\mu$Jy/beam with a beam size of 35.0~$\times$~34.2~mas at a PA of 67.5$^{\circ}$. The black star in panel~(c) indicates the center of ellipse isophoto fitting. (d): Azimuthally averaged radial profile in panel~(a). The synthesized image was deprojected with PA of 156.3$\degr$ and $i$ of 36.1$\degr$, derived in our visibility analyses in \S~\ref{subsec:mcmc}. The r.m.s.\ noise of the deprojected image is 9.8~$\mu$Jy/beam. (e): Azimuthal profile in the outer ring at $r\sim$20~au in the deprojected image. Two prominent asymmetries at PA of $\sim$270\degr\ and $\sim$180\degr\ in Figure~\ref{figA9} are labeled as blobs~A and B.} \label{fig1}
\end{figure}

The final synthesized dust continuum image of combined both data is shown in Figure~\ref{fig1}. In the \verb#CLEAN# task, we set the $uv$-taper to obtain a nearly circular beam pattern (Table~\ref{tab:table3}), and we do not use the `multi-scale' option. The r.m.s.\ noise in the region far from the object is 8.4~$\mu$Jy/beam with a beam size of 35.0~$\times$~34.2~mas at a position angle (PA) of 67.5$\degr$. 

The $^{12}$CO~$J=2\rightarrow1$ line data in both the long and short baseline data were extracted by subtracting the continuum in the $uv$ plane with the \verb#uvcontsub# task in the CASA tools. The combined line image cube with channel widths of 0.7~km/s was produced by the \verb#CLEAN# task. We also set the $uv$-taper to obtain the nearly circular beam pattern (Table~\ref{tab:table3}). The integrated line flux map (moment~0) and the intensity-weighted velocity map (moment~1) are shown in Figure~\ref{fig4} while channel maps at $-$1.0 to $+$12.3~km/s are shown in Figure~\ref{figA3} in the Appendix. The r.m.s.\ noise in the moment~0 map is 3.0~mJy/beam$\cdot$km/s with a beam size of 45.6~$\times$~45.4~mas at a PA of $-$77.9$^{\circ}$ while that in the moment~1 map at the 0.7~km/s bin is 589~mJy/beam. The peak SN ratio is 15.9 in the channel map of $+$2.5~km/s.

\begin{deluxetable*}{lll}
\tablewidth{0pt} 
\tablenum{1}
\tablecaption{ALMA Band 6 Observations and Imaging Parameters\label{tab:table3}}
\tablehead{
\colhead{} & \colhead{Long Baseline} & \colhead{Short Baseline}  
}
\startdata
Observing date (UT)             & 2019.Jun.05                & 2015.Aug.12        \\
Configuration                   & C43-9/10                   & --- \\
Project code                    & 2018.1.01755.S             & 2013.1.00498.S \\
Time on source (min)            & 134.3                      & 14.2 \\
Number of antennas              & 44                         & 44 \\
Baseline length                 & 83.1~m to 15.2~km          & 15.1~m to 1.6~km\\
Baseband freq.\ (GHz)           & 213.5, 216.3, 228.0, 230.0 & 217.0, 218.8, 219.3, 219.7, \\
                                &                            & 220.2, 230.7, 231.2, 232.3 \\
Channel width   (MHz)           & 15.63, 15.63, 15.63, 0.488 & 15.63, 7.813, 7.813, 0.488, \\
                                &                            & 0.488, 0.244, 3.906, 15.63 \\ 
Continuum band width (GHz)      & 7.5                        &  6.56 \\
Bandpass calibrator             & J0423$-$0120               &  J0423$-$0120 \\
Flux calibrator                 & J0423$-$0120               &  J0510$+$1800 \\
Phase calibrator                & J0431$+$1731               &  J0510$+$1800 \\
\hline \hline
                                & \multicolumn{1}{c}{Dust continuum} & \multicolumn{1}{c}{$^{12}$CO~$J=2\rightarrow1$} \\
\hline
Robust clean parameter          & 0.7                        & 2.0 \\
$uv$-taper Gaussian parameter   & 3.5~$\times$~50.0~M$\lambda$ at PA of 115$\degr$ & 3.4~$\times$~50.0~M$\lambda$ at PA of 120$\degr$\\
Beam shape                      & 35.0~$\times$~34.2~mas at PA of 67.5$\degr$ & 45.6~$\times$~45.4~mas at PA of $-$77.9$\degr$ \\
r.m.s.\  noise ($\mu$Jy/beam)   & 8.4                        & 2988 (moment 0) \\
                                &                            & 589 (moment 1 at 0.7~km/s bin) 
\enddata
\end{deluxetable*}

\begin{figure}[ht!]
\begin{centering}
\includegraphics[clip,width=\linewidth]{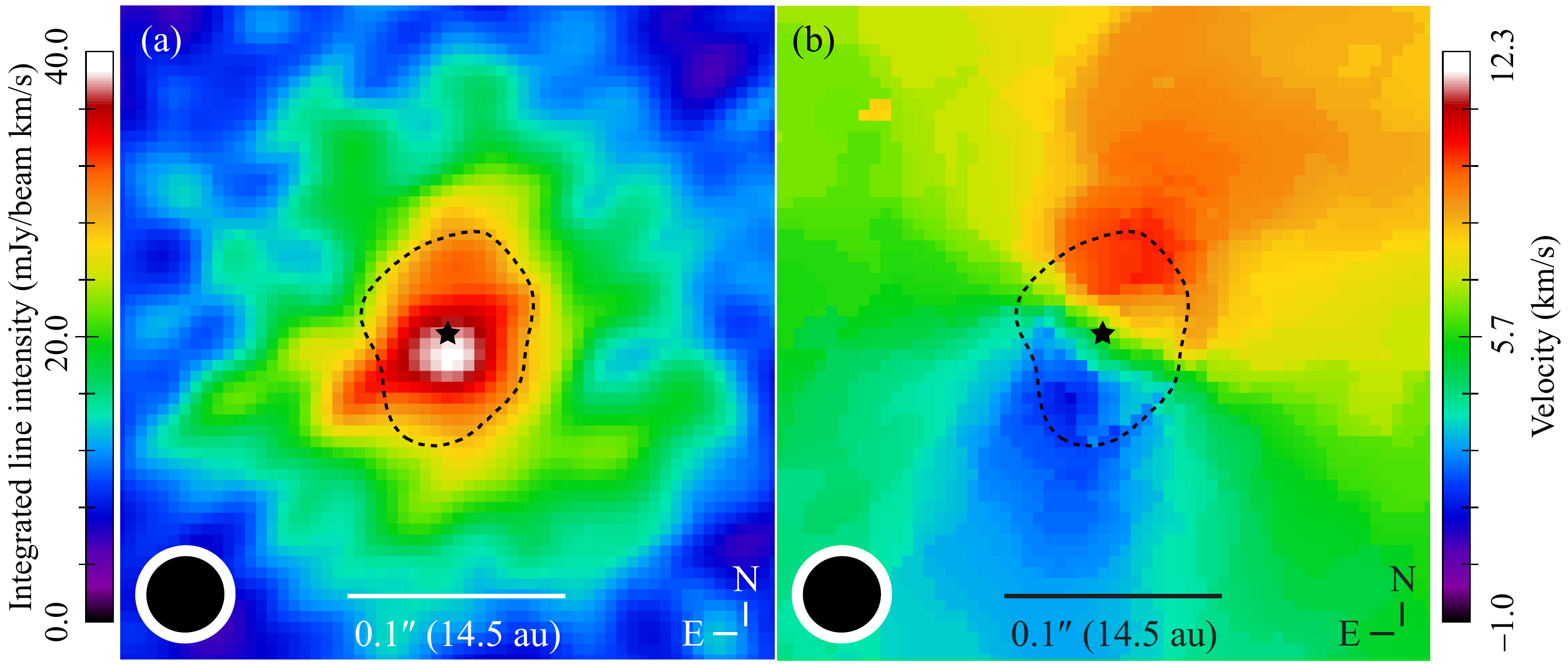}
\end{centering}
\caption{Integrated line flux map (a: moment~0) and intensity-weighted velocity map (b: moment~1) of $^{12}$CO~$J=2\rightarrow1$. The r.m.s.\ noise in the moment~0 map is 3.0~mJy/beam$\cdot$km/s with a beam size (white ellipse in the lower left corner) of 45.6~$\times$~45.4~mas at a PA of $-$77.9$^{\circ}$. Black dotted line represents the 15~$\sigma$ contour of the inner ring in the dust continuum image (1~$\sigma$ of 8.4~$\mu$Jy/beam; Figure~\ref{fig1}), and its beam size of 35.0~$\times$~34.2~mas at a PA of 67.5$^{\circ}$ is shown as block ellipse in the lower left corner. The positional error in the peak position is assumed to be the values of the beam size divided by the SN ratio, i.e., $\sim$5~mas. The black stars in both panels indicate the center of ellipse isophoto fitting in the dust continuum image.
}\label{fig4}
\end{figure}

\section{Results} \label{sec:result}

Figure~\ref{fig1} shows the dust continuum images of the DM~Tau disk combining long and short baseline data at band 6. The entire disk is shown in Figure~\ref{fig1}(a). As reported by \citet{kudo2018}, we confirmed three components in the disk: the inner ring at $r\lesssim$10~au, the outer ring at $r\sim$20~au, and the extended structure at $r\gtrsim$60~au, as noted in Figure~\ref{fig1}(d). We discovered that the extended structure consists of two faint rings at $r\sim$ 90 and $\sim$110~au with SN ratios of $\sim$6 and $\sim$5, respectively. This ``w"-shaped double gap structure, found in a number of other disks, could be produced by a super-Earth mass planet in a low viscosity environment \citep{dong17doublegap,dong18doublegap,pere2019a,facc2020a}. 

The structure of the outer ring at an SN of $\sim$70 is consistent with that in \citet{kudo2018}. In the image subtracting the 180\degr-rotated image in Figure~\ref{figA9}, we found two prominent asymmetries at PA of $\sim$270\degr\ and $\sim$180\degr\ (hereafter blobs~A and B) in the outer ring. These two blobs can be seen in the dust continuum image (Figure~\ref{fig1}b) and the azimuthal profile of the outer ring at $r\sim$20~au in Figure~\ref{fig1}(e), i.e., these two locate at the same radial location. We also synthesized the dust continuum images with different imaging parameters (e.g., using the multi-scale option) in the CLEAN task in Appendix~\ref{secA:multi} and Figure~\ref{figA14}, and confirmed the presence of the two blobs at roughly the same location in the images with different parameters. The contrasts of these two relative to the opposite side of the ring are $\sim$1.1$\times$ to $\sim$1.2$\times$. Note that blob~A has already been reported by \citet{kudo2018}. The total flux density derived by visibility fitting in \S~\ref{subsec:mcmc} is 94.74~$\pm$~9.47~mJy with assuming a 10~\% uncertainty in absolute flux calibration, consistent with previous single-dish observations (109~$\pm$~13~mJy, \citealp{beck90}) and previous ALMA observations (93.3~$\pm$~0.5~mJy by visibility fitting in \citealp{kudo2018}). The peak brightness temperature in the outer ring except blobs (i.e., at the inner edge of the outer ring, which is calculated from the best-fit modeled image in visibility fitting (\S~\ref{subsec:mcmc}), is 17.1~$\pm$~1.2~K with assuming a 10~\% uncertainty in absolute flux calibration. The optical depth $\tau_{\nu}$ is calculated with the relationship:
\begin{eqnarray}
I_{\nu} = B_{\nu}\,T_{\rm mid}\{1-{\exp}(-\tau_{\nu})\},
\end{eqnarray}
where $I_{\nu}$, $B_{\nu}$, and $T_{\rm mid}$ are intensity, the Plunck function, and the midplane temperature, respectively. We use the midplane temperature profile with the simplified expression for a passively heated, flared disk in radiative equilibrium \citep[e.g.,][]{dull2001}:
\begin{eqnarray}
T_{\rm mid}(r) = \left(\frac{\phi L_{*}}{8\pi r^{2}\sigma_{\rm SB}}\right)^{0.25},
\end{eqnarray}
with $L_{*}$ the stellar luminosity (taken as 0.36~$L_{\odot}$ from \citealp{mana14}), $\phi$ the flaring angle (taken as 0.02), and $\sigma_{\rm SB}$ the Stefan-Boltzmann constant. At the outer ring at $r=$20~au, $T_{\rm mid}$ is estimated to 21.6~K, and thus, the optical depth $\tau_{\nu}$ is calculated to 1.3$^{+0.3}_{-0.2}$.

Since measuring the flux density of the inner ring in the synthesized image is not straightforward, potentially due to a contamination of the bright outer ring, we derive it in the best-fit modeled image in the visibility fitting (\S~\ref{subsec:mcmc}). The peak brightness temperature and the flux density of the inner ring within the cavity of the outer ring calculated with the best-fit modeled image are 11.0$^{+0.6}_{-0.5}$~K and 1.74~$\pm$~0.17~mJy, respectively, with assuming a 10~\% uncertainty in absolute flux calibration. The flux density is similar to that in previous studies \citep[1.33~mJy in][]{kudo2018}, and is converted to a total mass (gas~$+$~dust) of 0.04~$M_{\rm Jup}$ assuming a distance of 145~pc, an opacity per unit dust mass $\kappa_{\nu}=$2.3~cm$^2$~g$^{-1}$ at 230~GHz \citep{beckwith+91}, a temperature of 100~K, and a gas-to-dust mass ratio of 100. Note that \citet{fran2020a} suggested the gas-to-dust mass ratio of 10$^{4}$~--~10$^{5}$ for the inner ring of DM~Tau assuming viscosity $\alpha$ of 10$^{-3}$, and thus, the inner ring may be two or three orders magnitude more massive.

\citet{kudo2018} reported a possible asymmetry in the inner ring, i.e., 20~\% brighter emission in the northwest. The inner ring in our new data shows that the southeast region is $\sim$20 \% ($\sim$3~$\sigma$) brighter than the northwest region (Figure~\ref{fig1}b). However, no such asymmetries in the inner ring can be seen in the image subtracting the 180\degr-rotated image in Figure~\ref{figA9} where only asymmetric signals are contained (see explanations of method~A in \S~\ref{sec:obs})\footnote{A demerit of this method is the fact that the noise level is $\sim$$\sqrt{2}$$\times$ higher than the normal dust continuum image because of imaging with only the imaginary part. Therefore, asymmetries with a small contrast can be elusive.}. Therefore, as these asymmetries could be the result of image reconstruction artifacts, more data are necessary to confirm both the asymmetries and morphological variability in the inner ring. 

The inner ring shown in Figure~\ref{fig1}(c) is likely to have a different PA than that of 157.8\degr\ in the DM~Tau system \citep{kudo2018}. The bright part in the south region in the inner ring is located at a PA of $\sim$180\degr. Since the beam shape is close to circular, the shape of the inner ring is unlikely to be affected by the beam elongation. We performed ellipse isophoto fitting of the inner ring at 8~$\sigma$ level, and found that the PA of the inner ring is 172.1\degr~$\pm$~4.2\degr. The difference is significant at 3.5~$\sigma$. Furthermore, \citet{fran2020a} found that the PA of the inner ring in \citet{kudo2018} is 141\degr~$\pm$~7\degr\ by Gaussian fitting in the image domain, which suggests that the PA of the inner ring in our new data varies comparing with previous observations in \citet{kudo2018}. These results motivated us to perform visibility analyses to test whether or not the PA of the inner ring is different from that of the system and previous observations, because there is a possibility of image reconstruction artifacts in the inner ring (see \S~\ref{subsec:mcmc}).

The integrated line flux map of $^{12}$CO~$J=2\rightarrow1$ in Figure~\ref{fig4}(a) shows a single-peak symmetric structure with a peak flux of 39.2~mJy/beam$\cdot$km/s at 13~$\sigma$, while \citet{kudo2018} noted that the peak emission at 9~$\sigma$ is shifted with $\sim$20~mas towards north\footnote{We found that errors in the $^{12}$CO moment~0 map estimated in \citet{kudo2018} are updated from 3.5 to 6.0~mJy/beam$\cdot$km/s.}. Assuming the positional errors are the values of the beam size divided by the SN ratio, the positional error of the peak $^{12}$CO emissions in \citet{kudo2018} is $\sim$8~mas, i.e., a $\sim$2.5~$\sigma$ deviation. These positional shifts between two epochs could also be artifacts. More data are needed to confirm or reject the time variability in the inner ring. The intensity-weighted velocity map is also shown in Figure~\ref{fig4}(b) and is consistent with that in \citet{kudo2018}.

\section{Modeling} \label{sec:model}
\subsection{Visibility fitting} \label{subsec:mcmc}

As the spatial scale of the inner ring is a few times the beam size, the structure in the inner ring corresponds to high spatial frequency components in the visibilities. In general, visibility data at a high spatial frequency is more sparse even in ALMA observations, potentially resulting in image reconstruction artifacts. To confirm the different PAs between the inner and outer rings inferred in \S~\ref{sec:result}, we performed forward modelling in which observed visibilities are reproduced with a parametric model of the disk by utilizing all spatial frequency information. 

\begin{figure}[ht!]
\begin{centering}
\includegraphics[clip,width=\linewidth]{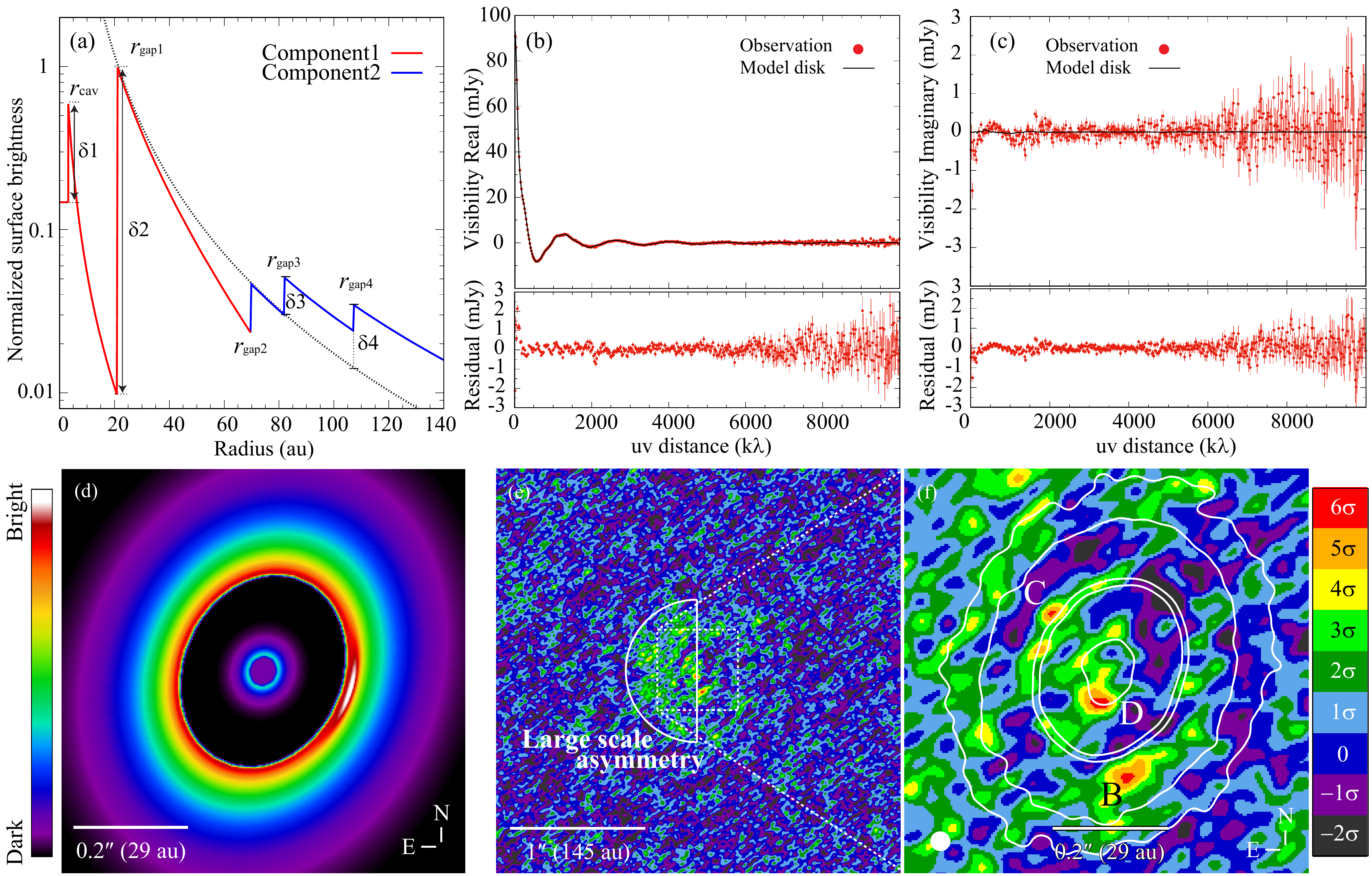}
\end{centering}
\caption{(a) Surface brightness profile for the best-fit model with notes regarding our visibility analyses. We divided the model surface brightness profile into two components: component~1 (red) for the inner and outer rings at $r\lesssim$ 60~au and component~2 (blue) for the extended structure at $r\gtrsim$ 60~au. The gap structures are created by scaling down/up the surface brightness by a factor $\delta$. These two components are normalized at $r=r_{\rm gap1}$ as shown by the black dotted line. To reproduce the nearly flat radial profile at $\sim$60~au~$<r<$~$\sim$80~au in Figure~\ref{fig1}(d), we added a pseudo ring in this region (i.e., at $r_{\rm gap_{2}} < r  < r_{\rm gap_{3}}$. (b and c) Real and imaginary parts of the visibilities for the observations (red dots) and the best-fit model (black line) in the top panel. The bottom panel shows residual visibilities between observations and the best-fit model. (d) Magnified image (0\farcs6~$\times$~0\farcs6) of the  best-fit model. (e and f) Residual image (3$\arcsec$~$\times$~3$\arcsec$) and its magnified image (0\farcs6~$\times$~0\farcs6). The large scale asymmetry in the east part of the disk is indicated by white solid line. The black dotted contours represent the dust continuum at SN ratios of 15 and 30. 
}\label{fig2}
\end{figure}

In the literature, the parametric disk model is often described with Gaussian rings \citep[e.g.,][]{zhan16,pini18a}. On the other hand, since the radial profile of the outer ring at $r=$20~au around DM~Tau shows an exponential profile (Figure~\ref{fig1}d), we describe the surface brightness distributions of the disk in our model with a simple power-law radial profile with an exponential taper at the outside:

\begin{eqnarray}
I(r) \propto \sum_{i=1}^{2} \alpha_{i} \left(\frac{r}{r_{c_{i}}}\right)^{-\gamma_{\rm{p},i}}{\rm exp}\left[-\left(\frac{r}{r_{c_{i}}}\right)^{\gamma_{\rm{e},i}}\right], 
\end{eqnarray}
where $\alpha_{i}$, $r_{c_{i}}$, $\gamma_{\rm{p},i}$, and $\gamma_{\rm{e},i}$ are a scaling factor, a characteristic scaling radius, and exponents of the power-law and the exponential taper, respectively. We divided the disk into two global components ($i$, Figure~\ref{fig2}a) because the inner and outer rings (component~1) are roughly one order of magnitude brighter than the extended outer structure (component~2). In the radial direction, we have the following scaling factors:

\begin{eqnarray*}
\alpha_{1} {\rm (component\ 1)} &=& \left \{
\begin{array}{llllllll}
\delta_{\rm {2}}                     &{\rm for}& r_{\rm cav}     &<& r  &<& r_{\rm gap_{1}}\\
1                                    &{\rm for}& r_{\rm gap_{1}} &<& r  &<& r_{\rm gap_{2}}\\
0                                    &{\rm for}& r_{\rm gap_{2}} &<& r  & & \\
\end{array}
\right. \\
\alpha_{2} {\rm (component\ 2)} &=& \left \{
\begin{array}{llllllll}
0                                    &{\rm for}&                 & & r  &<& r_{\rm gap_{2}}\\
1                                    &{\rm for}& r_{\rm gap_{2}} &<& r  &<& r_{\rm gap_{3}}\\
\delta_{\rm {3}}                     &{\rm for}& r_{\rm gap_{3}} &<& r  &<& r_{\rm gap_{4}}\\
\delta_{\rm {4}}                     &{\rm for}& r_{\rm gap_{4}} &<& r. & &
\end{array}
\right.
\end{eqnarray*} 
At $r < r_{\rm cav}$, we set a constant value with a depletion factor ($\delta_{1}$) relative to the brightness at $r = r_{\rm cav}$. We note that since the radial profile at $\sim$60~au~$<r<$~$\sim$80~au is likely to be flat (Figure~\ref{fig1}d), we added a pseudo ring in this region (i.e., at $r_{\rm gap_{2}} < r  < r_{\rm gap_{3}}$ in Figure~\ref{fig2}a) to reproduce the nearly flat radial profile. Two components are normalized at $r=r_{\rm gap_{1}}$ (Figure~\ref{fig2}a). The total flux density ($F_{\rm total}$) is also set as a free parameter. 
The disk inclination ($i$) and PA in the inner ring and the system (meaning the outer ring $+$ the extended structure hereafter) are set as free parameters, i.e., $i_{\rm inner\ ring}$, PA$_{\rm inner\ ring}$, $i_{\rm system}$, and PA$_{\rm system}$. 
We fix the phase center. 

In addition to the above disk, we also add the model of blob~A at PA of $\sim$270\degr\ (Figure~\ref{figA9}) defined as the elliptical gaussian function in the polar coordinate as follows:

\begin{eqnarray*}
{\rm A} &=& \frac{{\rm cos}^{2}\,{\rm PA_{blob}}}{\sigma_{\theta}^{2}} + \frac{{\rm sin}^{2}\,{\rm PA_{blob}}}{\sigma_{r}^{2}}, \nonumber \\
{\rm B} &=& 2\left(\frac{1}{\sigma_{\theta}^{2}} - \frac{1}{\sigma_{r}^{2}}\right)\, {\rm sin}\,{\rm PA_{blob}}\ {\rm cos}\,{\rm PA_{blob}}, \\ 
{\rm C} &=& \frac{{\rm sin}^{2}\,{\rm PA_{blob}}}{\sigma_{\theta}^{2}} + \frac{{\rm cos}^{2}\,{\rm PA_{blob}}}{\sigma_{r}^{2}}, \\
{\rm Z} &=& {\rm A} (\theta - \theta_{\rm blob})^{2} + {\rm B} (\theta - \theta_{\rm blob})(r - r_{\rm blob}) + {\rm C} (r - r_{\rm blob})^{2}, 
\end{eqnarray*}
\begin{equation}
I_{\rm blob}(r,\theta) \propto {\rm exp} \left(-\frac{1}{2} {\rm Z} \right),
\end{equation}
where PA$_{\rm blob}$ is PA of the major axis of the elliptical gaussian function in the polar coordinate, $r_{\rm blob}$ and $\theta_{\rm blob}$ are the radial and the azimuthal distances of blob~A in the polar coordinate, and $\sigma_{\theta}$ and $\sigma_{r}$ are standard deviations along the azimuthal and radial directions in the elliptical gaussian function, respcetively. The value of PA$_{\rm blob}$ is set to zero. The model image is finally rotated and magnified with PA$_{\rm system}$ and $i_{\rm system}$, respectively. The total flux of blob~A is normalized to $F_{\rm blob}$. Note that the values of ${\rm FWHM}_{\rm r,blob}$ and ${\rm FWHM}_{\rm \theta,blob}$ are equal to $2.355 \sigma_{r}$ and $2.355 \sigma_{\theta}$, respectively. Figure~\ref{figA11} shows the model image of blob~A. Note that since blob~B has lower brightness, we do not include blob~B in our model. In total, there are 25 free parameters in our model ($F_{\rm total}$, $F_{\rm blob}$, $r_{\rm cav}$, $r_{\rm gap_{1}}$, $r_{\rm gap_{2}}$, $r_{\rm gap_{3}}$, $r_{\rm gap_{4}}$, $\delta_{1}$, $\delta_{2}$, $\delta_{3}$, $\delta_{4}$, $\gamma_{\rm{p},1}$, $\gamma_{\rm{e},1}$, $\gamma_{\rm{p},2}$, $\gamma_{\rm{e},2}$, $r_{c_{1}}$, $r_{c_{2}}$, $i_{\rm inner\ ring}$, PA$_{\rm inner\ ring}$, $i_{\rm system}$, PA$_{\rm system}$, $r_{\rm blob}$, $\theta_{\rm blob}$, ${\rm FWHM}_{\rm r,blob}$, ${\rm FWHM}_{\rm \theta,blob}$).

The modeled disk image was converted to complex visibilities with the public python code \verb#vis_sample# \citep{loomis+17}, in which modeled visibilities are samples with the same ($u$, $v$) grid points with observations. The modeled visibilities are deprojected\footnote{Visibilities are deprojected in the $uv$-plane with the following equations \citep[e.g.,][]{zhan16}: 
 $u'    =   (u\,{\rm cos}\,{\rm PA_{system}} - v\,{\rm sin}\,{\rm PA_{system}}) \times {\rm cos}\,i_{\rm system},
  v'    =   (u\,{\rm sin}\,{\rm PA_{system}} - v\,{\rm cos}\,{\rm PA_{system}}),$
where $i_{\rm system}$ and ${\rm PA_{system}}$ are free parameters in our visibility analyses in \S~\ref{subsec:mcmc}.} with the system PA and $i$ as free parameters. The fitting is performed with a Markov chain Monte Carlo (MCMC) method in the \verb#emcee# package \citep{foreman-mackey+2013}. The log-likelihood function ln$L$ in MCMC fitting is 
\begin{eqnarray*}
{\rm ln}L = -0.5 \sum \left[f\,W_{j}\{({\rm Re}V_{j}^{\rm obs} - {\rm Re}V_{j}^{\rm model})^{2} + ({\rm Im}V_{j}^{\rm obs} - {\rm Im}V_{j}^{\rm model})^{2}\}\right],
\end{eqnarray*}
where  the subscript $j$ represents the $j$-th data. $V_{j}^{\rm obs}$, $V_{j}^{\rm model}$, and $W_{j}$ are observed and modeled visibilities, and weights, respectively. The value of $f$ is a factor between weights and standard deviations in the visibilities. To estimate the value of $f$, we calculate the standard deviations of the real and imaginary parts in 3~k$\lambda$ bin along the azimuthal direction in the visibility domain. The visibilities were deprojected with $i_{\rm system} =$36.1\degr\ and ${\rm PA_{system}} =$156.3\degr. Figure~\ref{figA8} shows the comparison between weights and standard deviations, and we found that the typical value of $f$ is 0.24. The weights are overestimated, or the noise is underestimated, vice versa. Our calculations used flat priors with the parameter ranges summarized in Table~\ref{tab:table1}. We ran 5000 steps with 100~walkers, and discarded the initial 500~steps as the burn-in phase based on the trace plot in Figure~\ref{figA7}.

The fitting results with errors computed from the 16th and 84th percentiles, the radial profile of best-fit surface brightness, the best-fit modeled image, and the probability distributions for the MCMC posteriors are shown in Table~\ref{tab:table1},  Figure~\ref{fig2}(a), Figure~\ref{fig2}(d), and Figure~\ref{figA2} in the Appendix, respectively. Though some parameters such as $r_{\rm blob}$ show double peaks in the probability distributions (Figure~\ref{figA2} in the Appendix), since the differences of double peaks are small, we only show the results for the best-fit model in Figure~\ref{fig2}. We subtracted modeled visibilities from observed ones, and made a CLEANed image (Figure~\ref{fig2}e and f). The reduced-$\chi^{2}$ is 1.7. We confirmed that the size of the inner cavity is 3~au in radius, which is consistent with the result in \citet{kudo2018}. Though Figure~\ref{fig1}(c) imply that the inner ring is misaligned to the outer ring, the values of PA and $i$ in the inner ring and the system are statistically same within 3~$\sigma$ in our visibility analyses.

\begin{figure}[ht!]
\begin{centering}
\includegraphics[clip,width=\linewidth]{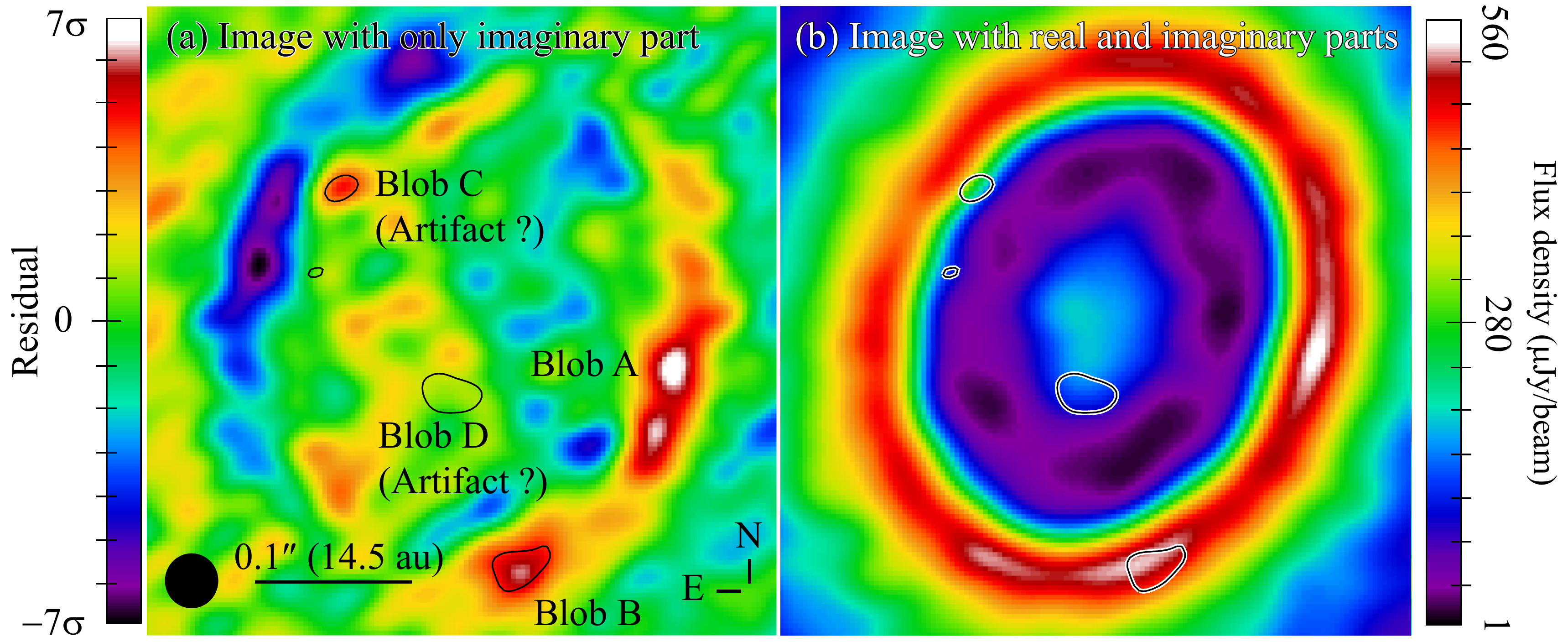}
\end{centering}
\caption{Left: The synthesized dust continuum image of short and long baseline data made with the only imaginary part (i.e., the image subtracting 180\degr-rotated image). Right: Normal dust continuum image. Black and white lines are 5~$\sigma$ contours in Figure~\ref{fig2}(f).
}\label{fig6}
\end{figure}

In the residual image in Figure~\ref{fig2}(f), we found additional two significant residual signals as labeling blob~C and D. Figure~\ref{fig6}(a) shows the dust continuum image with subtracting the 180\degr-rotated image, overlaying the contours of blob~B to D in Figure~\ref{fig2}(f). Though the counterparts of blob~B and C can be seen in Figure~\ref{fig6}(a), blob~C disappears in the image with different image shifts in Figure~\ref{figA9}, e.g., the image with $\Delta$R.A.=1~mas and $\Delta$DEC=1~mas. Hence we consider that blob~B is a real structure while blob~C might be an artifact. Furthermore blob~D has no
counterpart in Figure~\ref{fig6}(a), and thus, we also consider that blob~D is an artifact. By ellipse gaussian fitting in the image domain, blob~B is spatially resolved with the size of 119~$\times$~78~mas (17.3~$\times$~11.3~au).

The residual image also suggests the large scale asymmetry in the east part of the disk (Figure~\ref{fig2}e). The residuals of the real and imaginary parts in Figure~\ref{fig2}(b) and (c) also suggest the deviations in the shorter baseline at the $uv$-distance of less than $\sim$500~k$\lambda$ (corresponding to the scale of $\sim$0\farcs4). To check this large scale asymmetry, we compare the image with subtracting the modeled disk (Figure~\ref{fig2}e) and the image with subtracting the 180\degr-rotated image (Figure~\ref{fig6}a) in Figure~\ref{figA12} in Appendix. We found that both images in Figure~\ref{figA12} show the large scale asymmetry in the east part of the disk. Such large scale asymmetries have been reported in other disk systems potentially due to the shadow effect \citep[e.g., Figure~4 in][]{facc2020a}. The large scale asymmetry around DM~Tau (Figure~\ref{fig2}e; to be investigated elsewhere) has clumpy structures at $\sim$3--4~$\sigma$ level, and could potentially induce artificial clumps in the ring of blobs~A and B. The spatial distribution of low SNR clumps caused by thermal noises is expected to be random. If blobs~A and B are indeed such artificial clumps, it is unlikely for them to reoccur in different observations. We re-imaged dust continuum data in cycle~5 \citep{kudo2018} and found that both blobs can be seen at roughly the same locations (Figure~\ref{figA14}; Appendix~\ref{secA:multi}). This suggests that blobs~A and B are unlikely artificial clumps caused by noises. 

We also found that the central region at $r\lesssim$3~au is not empty because the value of $\delta_{1}=$0.24$^{+0.03}_{-0.05}$ is not zero at 4.8~$\sigma$. The total flux at $r\lesssim$3~au is roughly 50~$\mu$Jy. This result indicates the existence of unresolved ring structure at $r\lesssim$3~au because the SED of DM~Tau suggests the (nearly) empty cavity around the central star.

\begin{splitdeluxetable*}{cccccccccBccccccccccBcccccc}
\tabletypesize{\footnotesize}
\tablewidth{0pt} 
\tablenum{2}
\tablecaption{Results of MCMC fitting and its parameter ranges \label{tab:table1}}
\tablehead{
\colhead{$r_{\rm cav}$} & \colhead{$r_{\rm gap_{1}}$} & \colhead{$r_{\rm gap_{2}}$} & \colhead{$r_{\rm gap_{3}}$} & \colhead{$r_{\rm gap_{4}}$} & \colhead{$\delta_{\rm 1}$} & \colhead{$\delta_{\rm 2}$} & \colhead{$\delta_{\rm 3}$} & \colhead{$\delta_{\rm 4}$} & \colhead{$\gamma_{\rm{p},1}$} & \colhead{$\gamma_{\rm{e},1}$} & \colhead{$\gamma_{\rm{p},2}$} & \colhead{$\gamma_{\rm{e},2}$} & \colhead{$r_{\rm c1}$} & \colhead{$r_{\rm c2}$} & \colhead{$i_{\rm inner\ ring}$} & \colhead{PA$_{\rm innner\ ring}$} & \colhead{$i_{\rm system}$} & \colhead{PA$_{\rm system}$} & \colhead{Flux$_{\rm total}$} & \colhead{Flux$_{\rm blob}$} & \colhead{$r_{\rm blob}$} & \colhead{$\theta_{\rm blob}$} & \colhead{${\rm FWHM}_{\rm r,blob}$} & \colhead{${\rm FWHM}_{\rm \theta,blob}$} \\
\colhead{(au)}              & \colhead{(au)}              & \colhead{(au)}          & \colhead{(au)}              & \colhead{(au)}          & \colhead{} & \colhead{}          & \colhead{}          & \colhead{}    & \colhead{}          & \colhead{}    & \colhead{}          & \colhead{}          & \colhead{(au)}         & \colhead{(au)}         & \colhead{($^{\circ}$)}         & \colhead{($^{\circ}$)} & \colhead{($^{\circ}$)}         & \colhead{($^{\circ}$)} & \colhead{(mJy)}   & \colhead{($\mu$Jy)} & \colhead{(au)} & \colhead{(\degr)}              & \colhead{(au)}              & \colhead{(\degr)}                     
}
\colnumbers
\startdata
3.22$^{+0.31}_{-0.12}$ &  21.25$^{+0.09}_{-0.03}$ &  69.24$^{+0.71}_{-1.42}$ &  83.07$^{+0.51}_{-0.45}$ & 111.51$^{+1.76}_{-1.40}$ & 0.24$^{+0.03}_{-0.05}$ & 0.012$^{+0.000}_{-0.000}$ & 2.08$^{+0.16}_{-0.13}$ & 3.40$^{+0.28}_{-0.23}$ & 1.87$^{+0.12}_{-0.11}$ & 1.63$^{+0.13}_{-0.19}$ & 2.30$^{+0.07}_{-0.11}$   & 1.04$^{+0.03}_{-0.03}$ & 45.88$^{+4.62}_{-5.65}$ & 118.72$^{+7.23}_{-21.26}$ & 26.67$^{+7.62}_{-1.41}$ & 157.26$^{+3.53}_{-2.60}$ & 36.11$^{+0.44}_{-0.23}$ & 156.31$^{+0.32}_{-0.48}$ & 94.74$^{+0.82}_{-1.74}$ & 222.0$^{+28.4}_{-10.6}$ &  23.65$^{+0.83}_{-1.64}$ &  9.37$^{+0.51}_{-0.45}$ &  1.71$^{+1.03}_{-0.49}$ & 45.38$^{+3.23}_{-2.81}$ \\
\{0.0 .. 5.8\} & \{14.5 .. 29.0\} & \{55.1 .. 72.5\} & \{72.5 .. 92.8\} & \{98.6 .. 118.9\} & \{0.0 .. 1.0\} & \{0.0 .. 1.0\} & \{1.0 .. 10.0\} & \{1.0 .. 10.0\} & \{-1 .. 3\} & \{-1 .. 3\} & \{0 .. 3\} & \{0 .. 3\} & \{1.45 .. 145.0\} & \{1.45 .. 145.0\} & \{25.0-75.0\} & \{140.0--180.0\} & \{25.0--45.0\} & \{140.0--180.0\} & \{90.0 .. 110.0\} & \{100.0 .. 500.0\} & \{14.5-29.0\} & \{0.0--15.0\} & \{0.0--5.8\} & \{0.0--100.0\} \\
\enddata
\tablecomments{
 Parentheses describe parameter ranges in our MCMC calculations. }
\end{splitdeluxetable*}

\subsection{SED fitting} \label{subsec:mcrt}

\begin{figure}[ht!]
\begin{centering}
\includegraphics[clip,width=\linewidth]{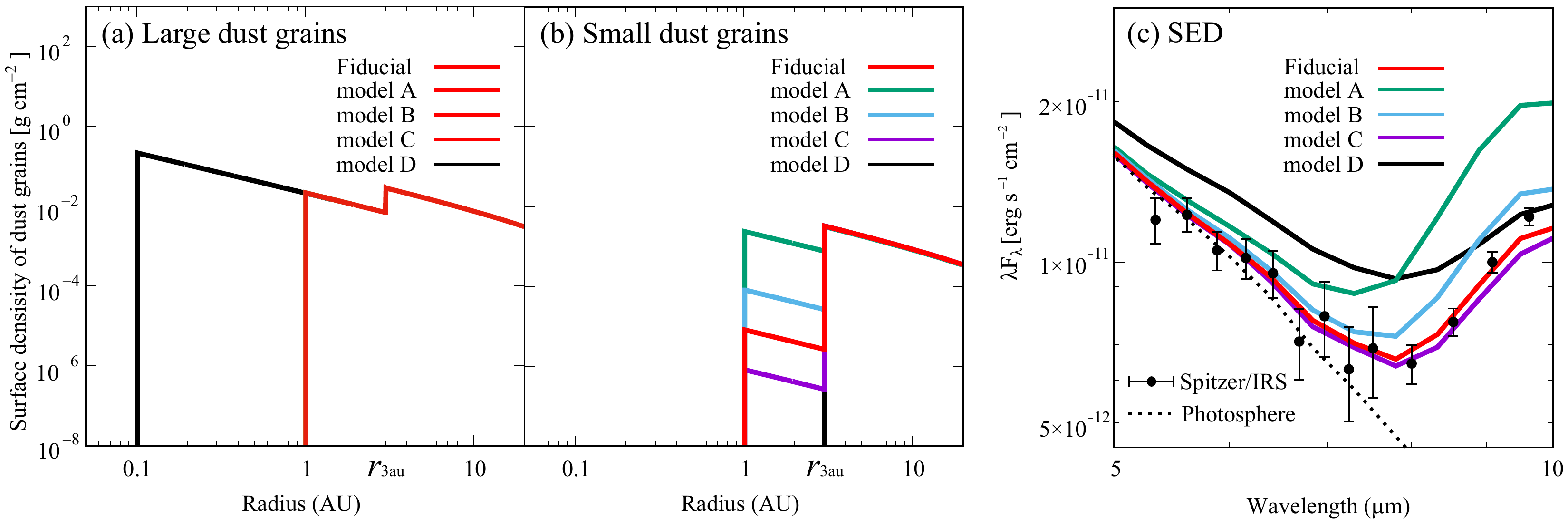}
\end{centering}
\caption{Results of MCRT modeling. We set $r_{\rm newcav}$ = 0.1 or 1.0~au while $r_{\rm 3au}=$ 3~au is derived from the visibility analyses in \S~\ref{subsec:mcmc}. (a and b) Surface density profiles of large (millimeter size) and small (sub-micron size) dust grains. The surface density at $r<$ 3~au is scaled down by a factor $\delta$. As DM~Tau SED has no NIR excess, we set no surface brightness at $r<r_{\rm newcav}$. (c) SEDs for our modeling. The black dots are DM~Tau spectra obtained with the Spitzer Space Telescope. 
}\label{fig5}
\end{figure}

The visibility analyses in \S~\ref{subsec:mcmc} suggests the significant millimeter emissions ($\sim$50~$\mu$Jy) in the central disk region within $r\lesssim$3~au. As DM~Tau has no or very little NIR excess in the SED \citep[e.g.,][]{calv05}, it has been believed that the dust grains inside the cavity are heavily depleted. The NIR excess mainly comes from small (sub-micron size) dust grains. To test the contributions of large (millimeter) dust grains in both the NIR excess in SED and the millimeter flux ($\sim$50~$\mu$Jy) of the central emission, we conducted radiative transfer modeling using a Monte Carlo radiative transfer (MCRT) code (\verb#HO-CHUNK3D#; \citealp{whit2013}). For this purpose, we put additional large and small dust grains inside the inner cavity at $r=$ 3~au. The new cavity radius at $r<$ 3~au is referred to as $r_{\rm newcav}$ hereafter. The fiducial surface density model and other models shown in Figure~\ref{fig5}(a). 

The MCRT code follows a two-layer disk model with small (up to micron size) dust grains in the upper disk atmosphere and large (up to millimeter size) dust grains in the disk mid-plane \citep[e.g.,][]{dale2006}. The modeled disk structure and dust properties in the MCRT code is described in our previous studies \citep{hashimoto+15}: small dust grains from the standard interstellar-medium dust model (a composition of silicates and graphites; a size distribution of $n(s) \propto s^{-3.5}$ from $s_{\rm min} = 0.0025$~$\mu$m to $s_{\rm max} = 0.2$~$\mu$m) in \citet{kim1994} and large dust grains (a composition of carbons and silicates; a size distribution of $n(s) \propto s^{-3.5}$ from $s_{\rm min} = 0.01$~$\mu$m to $s_{\rm max} = 1000$~$\mu$m) from Model~2 in \citet{wood2002}. The radial surface density is assumed to be a simple power-law radial profile similar to eq.~(3):  

\begin{eqnarray*}\label{eq1}
\Sigma(r) &=& \alpha \Sigma_{0} \left(\frac{r}{r_{c}}\right)^{-q}{\rm exp}\left[-\left(\frac{r}{r_{c}}\right)^{2-q}\right], \\
\alpha &=& \left \{
\begin{array}{llrlllll}
0                          &{\rm for}&                & & r  &<& r_{\rm newcav}\\
\delta                     &{\rm for}& r_{\rm newcav} &<& r  &<& {\rm 3~au}\\
1                          &{\rm for}& {\rm 3~au}     &<& r  &<& {\rm 20~au}
\end{array}
\right. 
\end{eqnarray*} 
where $\Sigma_{0}$ is the normalized surface density determined from the total (gas~$+$~dust) disk mass ($M_{\rm disk}$) assuming a gas-to-dust mass ratio of 100, $r_{c}$ is the characteristic radius of 50~au, $q$ is the radial gradient parameter, and $\alpha$ is the scaling factor for the surface density. As the main purpose of our MCRT modeling effort is to reproduce the DM~Tau SED at $\lambda \lesssim$10~$\mu$m, we set a grid size of $r=$ 20~au in the code, i.e., inside the 20~au cavity. We set $M_{\rm disk} =$ 0.1~$M_{\rm Jup}$ to reproduce a flux of 1.74~mJy inside the 20-au cavity at 1.3~mm, and fix $q=$ 1. The scale heights ($h$) of large and small grains are assumed to vary as a power law with a radius, i.e., $h \propto r^{p}$. To simplify, we assume $p=$ 1.25 with a typical midplane temperature profile of $T \propto r^{-0.5}$. We fix the scale heights of 1 and 4~au at $r=$ 100~au for large ($h^{\rm large}_{r=100{\rm au}}$) and small ($h^{\rm small}_{r=100{\rm au}}$) dust disks, respectively, taken from \citet{andr11} for the small dust disk. The mass fraction ($f$) of large dust grains in the total mass of dust grains is set to 0.9. The disk inclination is set to 30$\degr$. The \verb#HO-CHUNK3D# code calculates the accretion luminosity at the star based on the mass accretion rate. Half of the flux is emitted as X-rays (which heat the disk) and half as stellar flux at a higher temperature. We set a mass accretion rate of $\dot{M} =$~6~$\times $10$^{-9}$~$M_{\odot}$/yr \citep{mana14}. In the code, we vary three parameters: $r_{\rm newcav}$ (same values for large and small dust disks), $\delta^{\rm large}$, and $\delta^{\rm small}$ (where the superscript represents large or small dust grains) as shown in Table~\ref{tab:table2}.

\begin{deluxetable*}{ccccccccccc}
\tablewidth{0pt} 
\tablenum{3}
\tablecaption{Parameters in our MCRT modeling \label{tab:table2}}
\tablehead{
\colhead{Model} & \colhead{$R_{\rm newcav}$} & \colhead{$\delta^{\rm large}$} & \colhead{$\delta^{\rm small}$} & \colhead{$r_{c}$}  & \colhead{$p$}    & \colhead{$q$} & \colhead{$h^{\rm large}_{r=100au}$} & \colhead{$h^{\rm small}_{r=100au}$} & \colhead{$M_{\rm disk}$} & \colhead{$\dot{M}$} \\
\colhead{} & \colhead{(au)}          & \colhead{}    & \colhead{} & \colhead{(au)}   & \colhead{}    & \colhead{} & \colhead{(au)}     & \colhead{(au)} & \colhead{($M_{\rm Jup}$)} & \colhead{($M_{\sun}$~yr$^{-1}$)}       
}
\colnumbers
\startdata
fiducial & 1.0 & 3$\times$10$^{-1}$ & 1$\times$10$^{-3}$ & 50 & 1.25 & 1 & 1 & 4 & 0.1 & 6~$\times$~10$^{-9}$\\
A        & 1.0 & 3$\times$10$^{-1}$ & 3$\times$10$^{-1}$ & 50 & 1.25 & 1 & 1 & 4 & 0.1 & 6~$\times$~10$^{-9}$\\
B        & 1.0 & 3$\times$10$^{-1}$ & 1$\times$10$^{-2}$ & 50 & 1.25 & 1 & 1 & 4 & 0.1 & 6~$\times$~10$^{-9}$\\
C        & 1.0 & 3$\times$10$^{-1}$ & 1$\times$10$^{-4}$ & 50 & 1.25 & 1 & 1 & 4 & 0.1 & 6~$\times$~10$^{-9}$\\
D        & 0.1 & 3$\times$10$^{-1}$ & 0                  & 50 & 1.25 & 1 & 1 & 4 & 0.1 & 6~$\times$~10$^{-9}$\\
\enddata
\end{deluxetable*}

Figure~\ref{fig5}(c) shows the SEDs for each model. Our fitting procedure includes three steps, as follows.
\begin{enumerate}
\item We first set $r_{\rm newcav}$ = 1~au with varying values of the depletion factor $\delta$ with the same values in large and small dust disks to reproduce the flux of the central unresolved ring structure ($\sim$50~$\mu$Jy) at band~6, and found that $\delta$ = 3$\times$10$^{-1}$ is a reasonable parameter (model~A). 
\item As the SED for model~A largely emits at $\sim$10~$\mu$m (Figure~\ref{fig5}c), we only vary $\delta$ of small dust grains at 10$^{-2}$ to 10$^{-4}$ (models~B and C and fiducial model). We found that the fiducial model well reproduces the DM~Tau SED at $\lambda \lesssim$10~$\mu$m.  
\item We also set $r_{\rm newcav}$ = 0.1~au (model~D). However, even for $\delta$ = 0 for small dust grains, this model largely emits in the NIR wavelengths.
\end{enumerate}
In summary, the fiducial model could reasonably account for both the DM~Tau SED and the flux of the central unresolved ring structure. This could be because the midplane temperature at $r>$ 1~au is too low to emit at NIR, as shown in Figure~\ref{figA6} in the Appendix. Our modeling suggests that small dust grains inside the 3-au cavity are depleted while large ones remain present.

In our two modeling efforts for visibility analyses (\S~\ref{subsec:mcmc}) and SED (\S~\ref{subsec:mcrt}) fitting, we assume independent modeled disks. To check the consistency of the surface brightness for the two models, we plot them at $r<$ 20~au in Figure~\ref{figA4} in the Appendix. We confirm that the two radial profiles in our modeling efforts are consistent with each other. 

\section{Discussions} \label{sec:discuss}

\subsection{Multiple blobs} 

Asymmetric structures referred to as blobs in this paper are interpreted as dust trapped in gas vortices \citep[e.g.,][]{raettig2015} or gas horseshoes \citep[e.g.,][]{ragusa2017}. For the latter, only one blob is expected at the edge of a cavity, and thus, DM~Tau's multiple blobs would not be this case. Furthermore, a gas horseshoe is expected at the edge of a cavity opened by massive companions (i.e., brown dwarfs). Previous near-infrared sparse aperture masking interferometry \citep{will16} and radial-velocity measurement \citep{nguyen2012} have ruled out the presence of such companions in the cavity, disfavoring the gas horseshoe origin of DM Tau's blobs.
Note that though \citet{vandermarel20a} mentioned the spiral structures as a third origin of asymmetries, since scattered light images of DM~Tau is not available, it is unclear whether the spiral structures are responsible for DM Tau's blobs.

Theoretical works of vortices \citep[e.g.,][]{ono2018a} predict that multiple small vortices at similar radial locations tend to merge into one large vortex within hundreds of orbits. The existence of multiple blobs would therefore indicate the youth of vortices. Another interpretation of multiple blobs is a destruction of a large-scale vortex due to the heavy-core instability \citep{chang2010a}, triggered by a close to unity dust-to-gas mass ratio in the core of a vortex. Recent numerical simulations by \citet{li2020a} show multiple small blobs in the ring after the destruction of a large vortex. As the orbital number (system age divided by Keplerian orbital period at the radial location of the blob) of DM~Tau's blob~A is more than 10$^{4}$, DM~Tau's blobs may be the outcome of vortex destruction. In this case, the dust-to-gas mass ratio is expected to be close to unity. For DM~Tau however, the azimuthally-averaged dust-to-gas mass ratio is estimated to be $\sim$0.01 in the outer ring at $r\sim$20~au (Francis et~al. in~perp.), disfavoring the scenario. A third possibility is that the azimuthal position of trapped dust depend on the dust size \citep{baruteau2016}: centimeter-sized dust grains are trapped ahead of the gas vortex center in the azimuthal direction while millimeter-size dust grains concentrate closer to the vortex center. To examine this possibility, multiple wavelength observations to measure the spectral index sensitive to the grain size are necessary. 

\begin{figure}[ht!]
\begin{centering}
\includegraphics[clip,width=\linewidth]{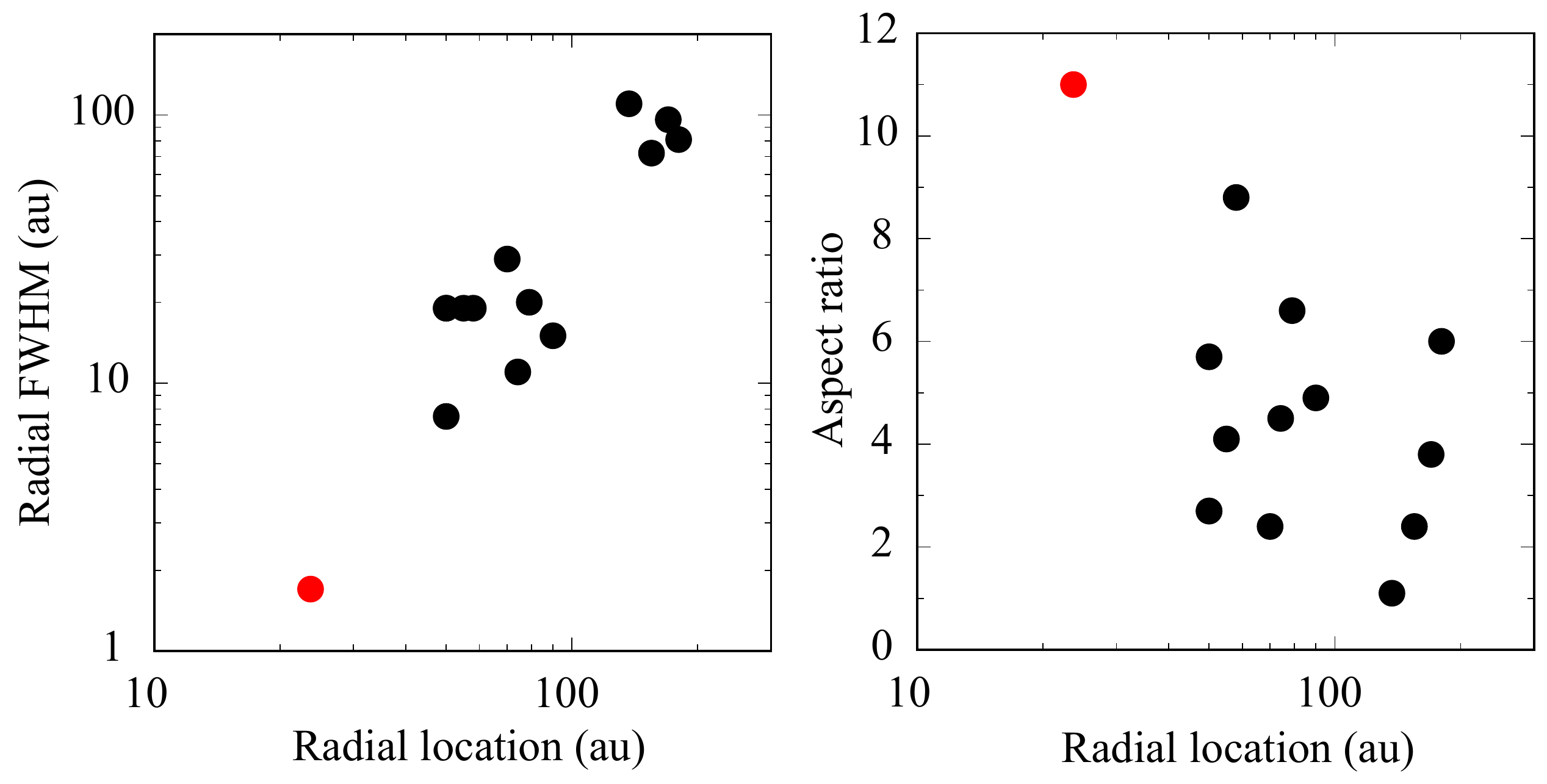}
\end{centering}
\caption{Comparison of the shape of DM~Tau's blob~A with other asymmetries. The aspect ratio is defined as the azimuthal width divided by the radial width. Red circle represent DM~Tau. Physical quantities are summarized in Table~\ref{tab:table4}.
}\label{fig7}
\end{figure}

We also compare the shape of DM~Tau's blob~A with other asymmetries. Asymmetric structures have been reported in roughly 10 protoplanetary disks \citep[e.g.,][]{fran2020a,vandermarel20a,tsukagoshi2019a,perez2018a,kraus+17}. Table~\ref{tab:table4} summarizes the physical quantities of blobs, mainly relevant to their morphology\footnote{V1247~Ori shows the crescent structure \citep{kraus+17}. However, we do not include  V1247~Ori because the structure was not characterized with the Gaussian profile \citep{kraus+17}. TW~Hya also shows the asymmetry \citep{tsukagoshi2019a}. Though all asymmetries in Table~\ref{tab:table4} are located at the inner/outer edges of the ring, TW~Hya's blob is not the case. TW~Hya's blob may be created by different mechanisms, and thus, we do not include TW~Hya's blob in  Table~\ref{tab:table4}.}. We found that DM~Tau's blob~A is located at the smallest radial location in the sample. Furthermore, its aspect ratio (the azimuthal width divided by radial width) is the largest. Figure~\ref{fig7} shows these observational results, which places DM~Tau's blob~A in a novel parameter space of asymmetries. 

\begin{deluxetable*}{cccccc}
\tablewidth{0pt} 
\tablenum{4}
\tablecaption{Physical quantities of blobs and other properties \label{tab:table4}}
\tablehead{
\colhead{Object} & \colhead{Radial location} & \colhead{Radial width} & \colhead{Azimuthal width} & \colhead{Aspect ratio} & \colhead{Refs} \\
\colhead{} & \colhead{(au)}        & \colhead{(au)}         & \colhead{(deg; au)}    &\colhead{}      & \colhead{} 
}
\colnumbers
\startdata
DM Tau     & 24  & 1.7  &  45; 18.7  & 11  & 1 \\
AB Aur     & 170 & 96   & 122; 361.8 & 3.8 & 2 \\
CQ Tau     & 50  & 19   & 59; 51.5   & 2.7 & 2 \\
           & 50  & 19   & 59; 51.5   & 2.7 & 2 \\
HD 34282   & 137 & 110  & 52; 124.3  & 1.1 & 3 \\
HD 34700   & 155 & 72   & 64; 173    & 2.4 & 4 \\
HD 142527  & 180 & 81   & 155; 486.7 & 6.0 & 2 \\
HD 143006  & 74.2  & 11 &38.4; 49.7  & 4.5 & 5 \\
IRS 48     & 70  & 29   & 58; 70.8   & 2.4 & 2 \\
MWC 758    & 50 & 7.5   & 49; 42.7   & 5.7 & 2 \\
           & 90  & 15   & 47; 73.8   & 4.9 & 2 \\
SAO 206462 & 79  & 20   & 96; 132.3  & 6.6 & 2 \\
SR 21      & 55  & 19   & 82; 78.7   & 4.1 & 2 \\
           & 58  & 19   & 165; 166.9 & 8.8 & 2 \\
\enddata
\tablecomments{
Radial width of IRS~48 is calculated with 2.17$\sigma_{r}$ because the radial profile of IRS~48 was found to be best fit with a 4th power in 2D Gaussian in \citet{vandermarel20a}. The aspect ratio is defined as the azimuthal width divided by the radial width. References for blob information of radial locations and shapes: 1) this work, 2) \citet{vandermarel20a}, 3) \citet{vanderpalas17}, 4) \citet{benac2020a}, 5) \citet{perez2018a}.
 }
\end{deluxetable*}

\subsection{Central emission} 
Our visibility analyses in \S~\ref{subsec:mcmc} suggest millimeter emissions (4.8~$\sigma$) within the 3~au cavity. Central point sources in the cavity of the disk can been seen in roughly half of the samples \citep[see Figure~1 in][]{fran2020a}. For these point sources, it has been shown that their total mm-dust mass is not correlated with the NIR excess, generally associated with small grains at the inner dust rim \citep{fran2020a}. Our modeling in \S~\ref{subsec:mcrt} supports the fact that even though DM~Tau has no NIR excess, there may still be a small ring at $\lesssim$3~au where large dust grains are dominant.

In the SED fitting results (\S~\ref{subsec:mcrt}), large dust grains in the 1-au ring are less depleted than small dust grains. One interpretation for this is collisional aggregation, i.e., grain growth. For sub-micron size dust grains, van der Waals forces cause the dust grains to stick when they meet each other by Brownian motion until they reach an upper size limit at which point they fragment to smaller dust grains through collisions. Numerical simulations by \citet{dull2005a} suggest a rapid depletion of small dust grains by grain growth on a timescale of less than 1~Myr for negligible particle fragmentation, and show a very weak NIR excess in the SED. This may be the case for DM~Tau's 1-au ring where grain growth efficiently occurs and the NIR excess is negligible. The collisional fragmentation could be suppressed in the 1-au ring around DM~Tau where dust grains could contain water ice, as discussed below \citep[e.g.,][]{wada2009}. We note that radial drift (\citealp{weid77,naka86}; in which particles embedded in a gaseous disk with a surface density that decreases outward feel a headwind, lose angular momentum to the gas, and drift toward the central star) may be invalid for millimeter dust grains at $r=$ 1~au while centimeter dust grains may efficiently drift there, as shown in \citet{brauer+2008}. 

The water snowline around typical T~Tauri stars is expected to be located at a few au from the central star \citep{notsu16}. Inside the snowline, the temperature exceeds the sublimation temperature ($T \sim$ 150~K; \citealp{notsu16}) and the water is released into the gas phase. Thus, dust grains inside the snowline contain rock and iron without water ice, and are believed to grow to form rocky planets. However, it has been believed that collisional fragmentations are dominant inside the water snowline due to a lack of water ice in the dust grains \citep[e.g.,][]{blum2008}. On the other hand, recent laboratory experiments by \citet{stei2019} show that silicate dust grains without water are an order of magnitude stickier than those with water. These results might invoke the possibility of the formation of rocky planetesimals.
In the case of DM~Tau, the disk midplane temperature is below 150~K except at the wall of the 1-au cavity (Figure~\ref{figA6} in Appendix), and thus, dust grains would contain water ice. This could be because the large depletion factor of 0.4 in the large dust grains in the 1-au ring results in a large optical depth, and the midplane temperature is under the conditions of radiative equilibrium. The flux of 50~$\mu$Jy in the 1-au ring translates to a total mass (gas~$+$~dust) of 0.4~$M_{\rm Earth}$ (a dust mass of 0.3~$M_{\rm Moon}$) assuming a distance of 145~pc, an opacity per unit dust mass $\kappa_{\nu}=$ 2.3~cm$^2$~g$^{-1}$ at 230~GHz \citep{beckwith+91}, a temperature of 100~K, and a gas-to-dust mass ratio of 100. We note that since dust grains in the 1-au ring might grow more than those in the outer disk region, as discussed in the previous paragraph, the dust opacity could be smaller than 2.3~cm$^2$~g$^{-1}$. Therefore, the derived dust mass of 0.3~$M_{\rm Moon}$ is probably a lower limit. These results suggest that small icy planets, i.e., mini Neptunes, may form in the 1-au ring around DM~Tau.

\section{Conclusion} \label{sec:conclusion}
We present new ALMA observations of DM~Tau including dust continuum images at 1.3~mm and $^{12}$CO~$J=2\rightarrow1$ emission maps. The dust continuum data with better sensitivity than previous studies in \citet{kudo2018} reveals (a) multiple ring structures at $r=$ 3, 20, 90, and 110~au, and (b) two blobs at PA of $\sim$180\degr\ and $\sim$270\degr\ in the outer ring at $r\sim$20~au. To characterize the inner ring regions at $r\lesssim$ 10~au, we analyze the dust continuum emissions in the visibility domain and conduct modeling efforts using the MCRT code \citep{whit2013}. Consequently, we found the unresolved 1-au ring inside the 3-au inner ring. Furthermore, model disks with different disk inclinations and PAs between the inner ring at $r\sim$3~au and the whole system suggests to that the inner ring is statistically aligned to the whole system within 3~$\sigma$.

DM~Tau's two blobs have the low contrast of $\sim$1.1$\times$ to $\sim$1.2$\times$ relative to the outer ring. These two blobs are located at the smallest radial location among 11 asymmetric disks. Furthermore, the aspect ratio of blob~A is the largest. These observational results, which places DM~Tau's blob~A in a novel parameter space of asymmetries. The origin of two blobs is not determined by our observations. The early phase of the vortex formation \citep[e.g.,][]{ono2018a}, the destruction of the large scale vortex \citep[e.g.,][]{li2020a}, or double continuum emission peaks with different dust sizes \citep{baruteau2016} could account for the multiple blobs. Future high spatial resolution observations in the multiple wavelengths with ALMA and JVLA would help identify the origin.

We also found the significant emissions with a lower limit mass of 0.4~$M_{\rm Earth}$ inside the 3~au cavity. As the DM~Tau SED shows negligible NIR excess, the inside of the 3-au cavity around DM~Tau has been believed to be a dust-free region. By fitting both the DM~Tau SED and the flux of the central emissions using the MCRT code, our modeling shows that inside the 3-au cavity, there is an additional 1-au dust ring where large (millimeter size) dust grains are less depleted than small (sub-micron size) dust grains. This would be due to efficient grain growth in the 1-au dust ring. Furthermore, our modeling indicates that the disk midplane temperature in the 1-au ring is less than the typical water sublimation temperature of 150~K \citep[e.g.,][]{notsu16}, which suggests that only small icy planets (i.e., mini Neptunes) could form even in the terrestrial planet formation regions around DM~Tau. 

\acknowledgments
We thanks the anonymous referee for a helpful review of the manuscript.
This paper makes use of the following ALMA data: ADS/JAO.ALMA\#2018.1.01755.S, ADS/JAO.ALMA\#2017.1.01460.S, and ADS/JAO.ALMA\#2013.1.00498.S. ALMA is a partnership of ESO (representing its member states), NSF (USA), and NINS (Japan), together with NRC (Canada), NSC (Taiwan), ASIAA (Taiwan), and KASI (Republic of Korea), in cooperation with the Republic of Chile. The Joint ALMA Observatory is operated by ESO, AUI/NRAO, and NAOJ.
This work is based in part on archival data obtained with the Spitzer Space Telescope, which was operated by the Jet Propulsion Laboratory, California Institute of Technology under a contract with NASA. Support for this work was provided by an award issued by JPL/Caltech.
This work was supported by JSPS KAKENHI Grant Numbers 19H00703, 19H05089, and 19K03932.
Y.H. is supported by the Jet Propulsion Laboratory, California Institute of Technology, under a contract with NASA. R.D. is supported by the Natural Sciences and Engineering Research Council of Canada through a Discovery Grant, and the Alfred P. Sloan Foundation through a Sloan Research Fellowship.

{\it Software}: \verb#vis_sample# \citep{loomis+17}, 
          \verb#HOCHUNK3D# \citep{whit2013}, 
          \verb#CASA# \citep{mcmu07}, 
          \verb#emcee# \citep{foreman-mackey+2013}

\appendix

\section{Dust continuum images synthesized with only the imaginary part}\label{secA:imag}

\begin{figure}[ht!]
\begin{centering}
\includegraphics[clip,width=\linewidth]{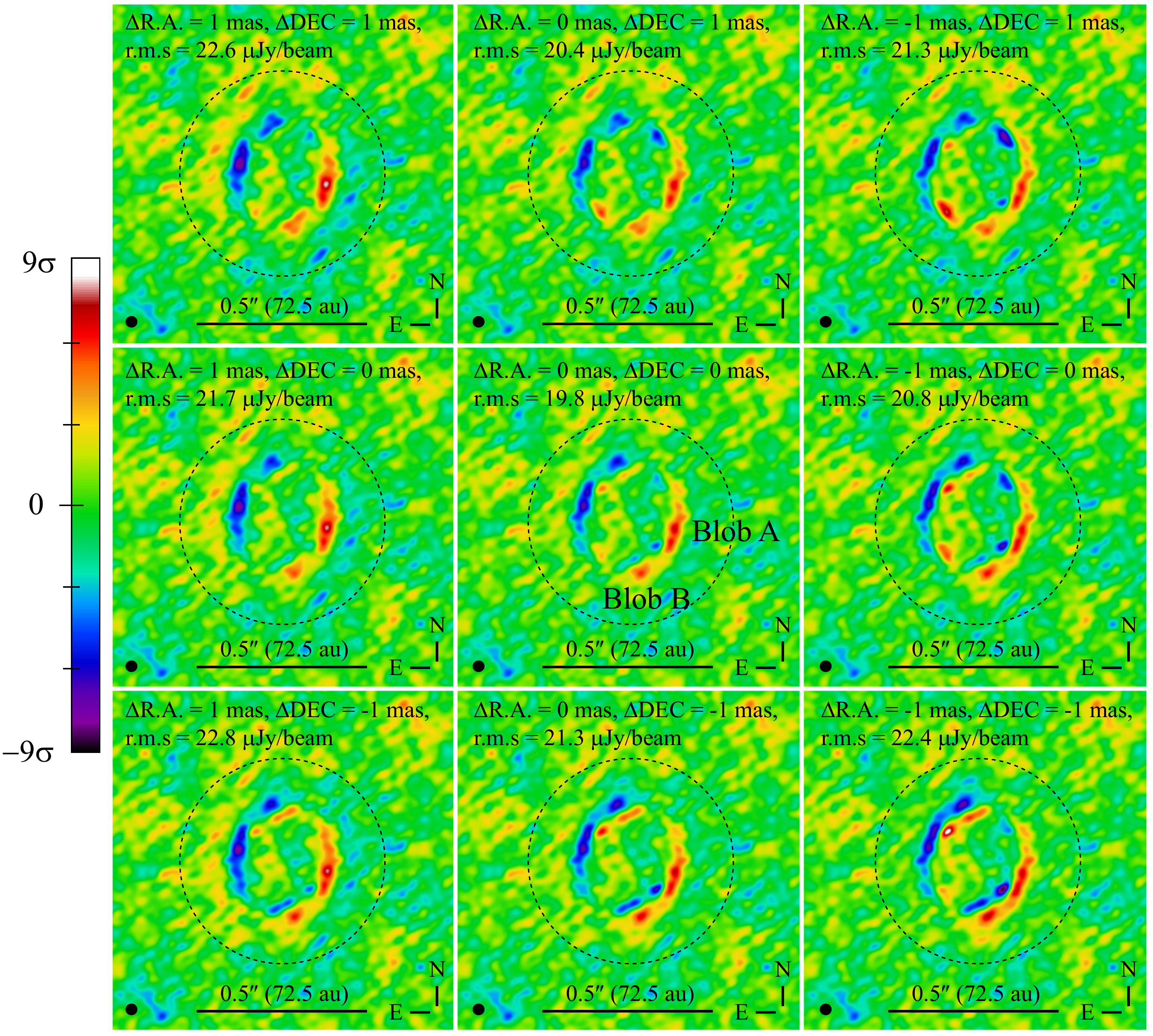}
\end{centering}
\caption{The dust continuum images synthesized with only the imaginary part of long baseline data by shifting 1~mas in R.A. and DEC directions. The r.m.s value is measured inside the black dotted circle with a radius of 0\farcs3. The 1~$\sigma$ noise is 11.6~$\mu$Jy/beam measured in the region far from the object.}\label{figA9}
\end{figure}

\begin{figure}[ht!]
\begin{centering}
\includegraphics[clip,width=\linewidth]{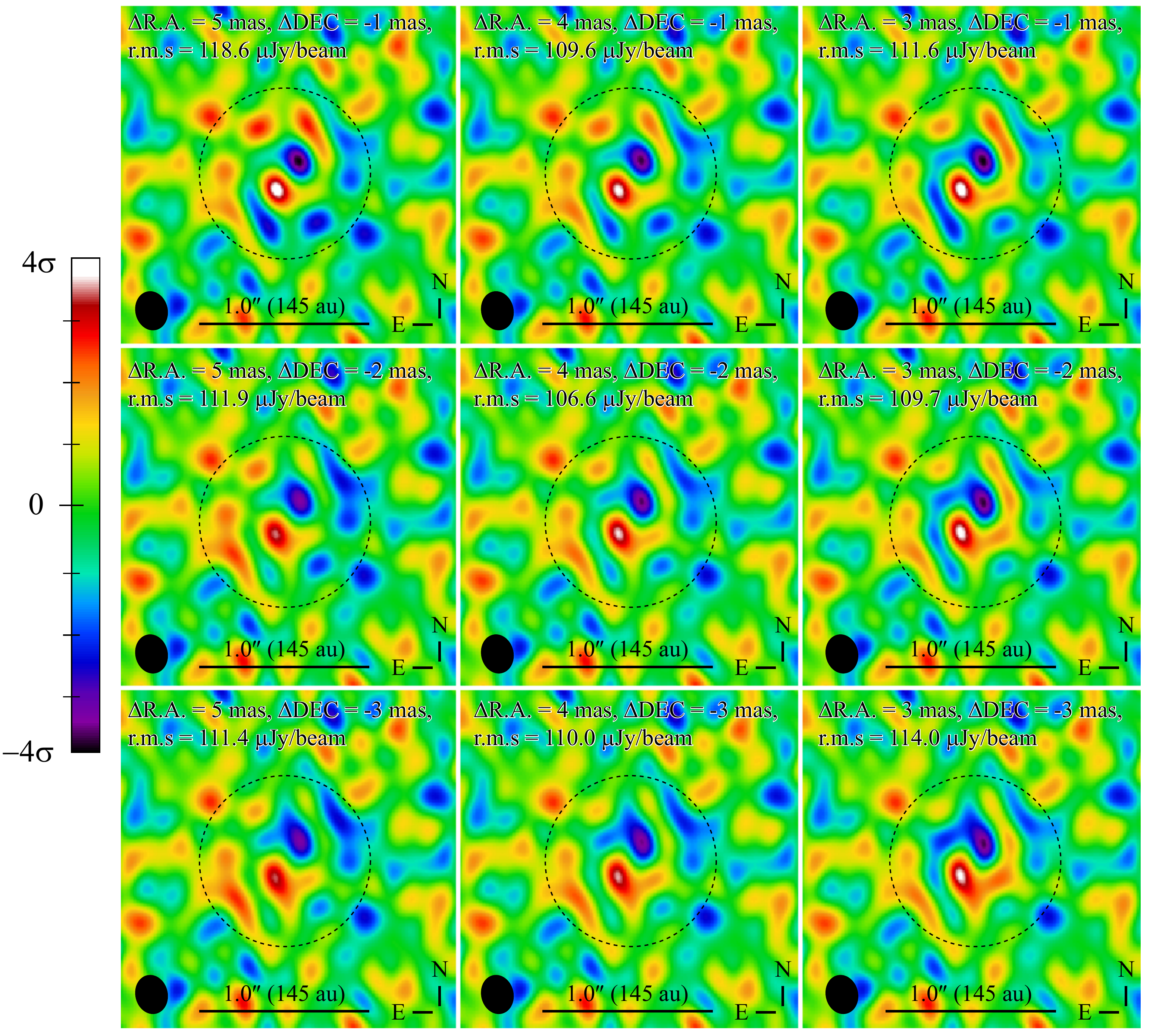}
\end{centering}
\caption{Same with Figure~\ref{figA9}, but short baseline data. The r.m.s value is measured inside the black dotted circle with a radius of 0\farcs5. The 1~$\sigma$ noise is 89.0~$\mu$Jy/beam measured in the region far from the object.}\label{figA10}
\end{figure}

Figure~\ref{figA9} and \ref{figA10} show the dust continuum images synthesized with only the imaginary part of long and short baseline data, respectively, by shifting 1~mas in R.A. and DEC directions. The image is shifted relative to the center of ellipse isophoto fitting (see \S~\ref{sec:obs}) in the visibility domain by the phase shift defined as $\exp\left[2\pi\left(u\,\Delta {\rm R.A.} +v\,\Delta {\rm DEC}\right)\right]$, where $u$ and $v$ are the spatial frequencies and $\Delta$R.A. and $\Delta$DEC are shift values, respectively. The r.m.s values are measured inside the black dotted circles with radii of 0\farcs3 and 0\farcs5 in Figure~\ref{figA9} and \ref{figA10}, respectively. The 1~$\sigma$ noises in Figure~\ref{figA9} and \ref{figA10} are 11.6~$\mu$Jy/beam and 89.0~$\mu$Jy/beam, respectively, measured in the region far from the object.

\section{Dust continuum images with different imaging parameters} \label{secA:multi}

\begin{figure}[ht!]
\begin{centering}
\includegraphics[clip,width=\linewidth]{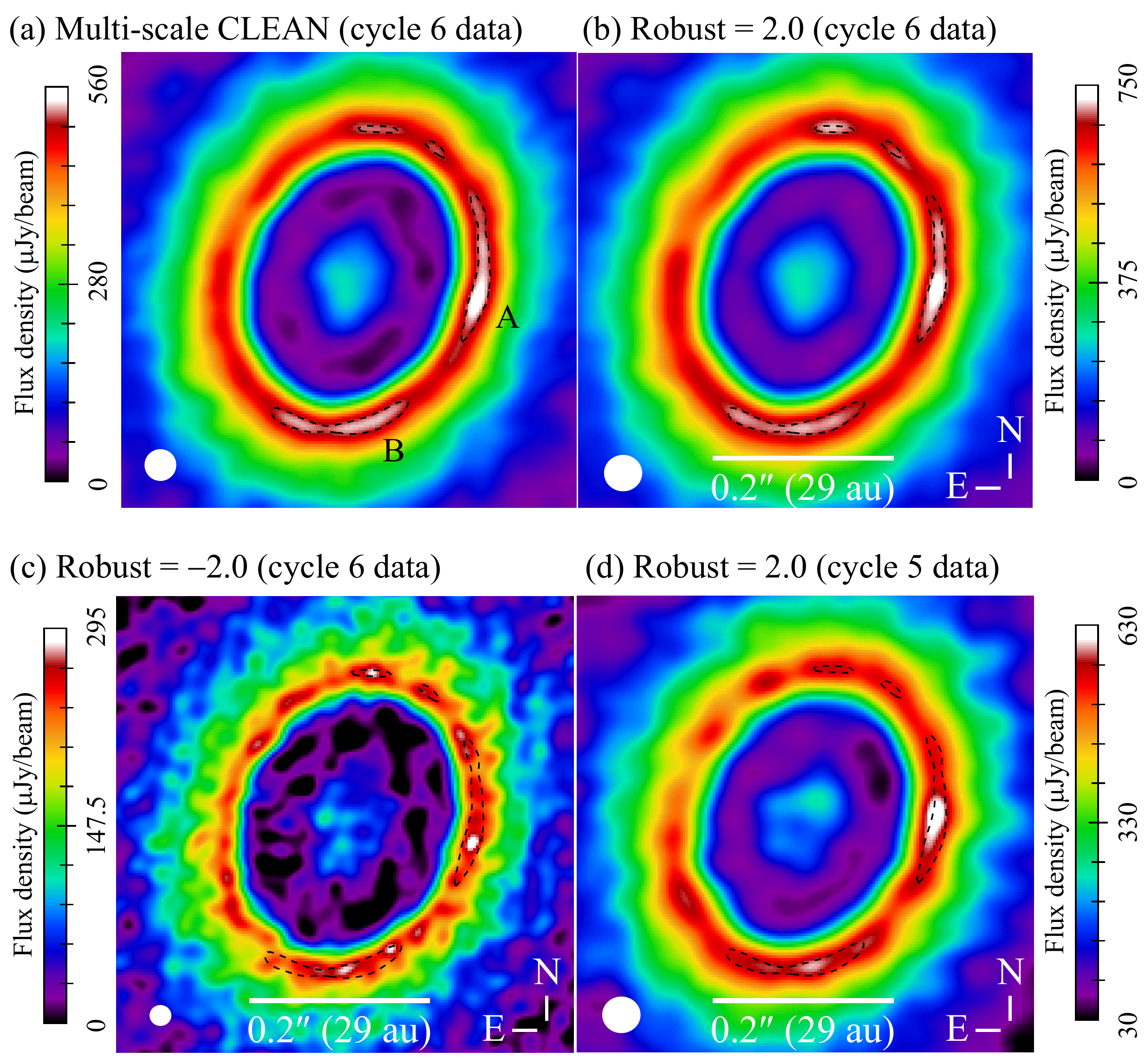}
\end{centering}
\caption{The dust continuum images with different imaging parameters of cycle~6 data in panels~(a) to (c) and cycle~5 data taken from \citet{kudo2018} in panel~(d).
(a) The dust continuum image synthesized with the multi-scale option and other parameters fixed as in Figure~\ref{fig1}. The image is consistent with Figure~\ref{fig1}. The noise level is 8.4~$\mu$Jy/beam with the beam shape of 35~$\times$~34~mas at PA of 67.5$\degr$. 
(b and c) Same with panel (a), but with robust=2.0 and $-$2.0 (the multi-scale option was not used). The noise levels and the beam shapes in panels (b) and (c) are 12.0~$\mu$Jy/beam with the beam of 42~$\times$~40~mas at PA of 46.8$\degr$ and 15.8~$\mu$Jy/beam with the beam of 24~$\times$~23~mas at PA of $-$12.9$\degr$, respectively. 
(d) Same with panel~(b), but cycle~5 data \citep{kudo2018}. The noise level is 11.8~$\mu$Jy/beam with the beam shape of 38~$\times$~37~mas at PA of 27.6$\degr$. The black dotted lines in all panels represent the 60~$\sigma$ contours in the dust continuum image in Figure~\ref{fig1}. 
}\label{figA14}
\end{figure}

We synthesized dust continuum images with different imaging parameters in the CLEAN task to examine the robustness of blobs~A and B. We also re-imaged the dust continuum data obtained in cycle~5 \citep{kudo2018} to check whether the two blobs are present in different datasets. To minimize the effect of the beam elongation, we adjusted the $uv$-taper parameters to obtain a nearly circular beam. Figure~\ref{figA14}(a) shows the dust continuum image using the same parameters with Figure~\ref{fig1}, but using the multi-scale option with scales of [0, 5, 10, 15, 25] in the CLEAN task. We confirmed that the image is consistent with Figure~\ref{fig1}. Figure~\ref{figA14}(b) and (c) are the same with Figure~\ref{figA14}(a), but with robust=2.0 and $-$2.0. Figure~\ref{figA14}(d) shows the dust continuum images obtained in cycle~5, with robust=2.0. Since we do not see significant differences in the images with or without the multi-scale option in Figure~\ref{figA14}(a) and Figure~\ref{fig1}, the multi-scale option was not used in Figure~\ref{figA14}(b to d). We confirmed the presence of blobs~A and B at roughly the same locations in all cases.

\section{$^{12}$CO~$J=2\rightarrow1$ channel maps}

\begin{figure}[ht!]
\begin{centering}
\includegraphics[clip,width=\linewidth]{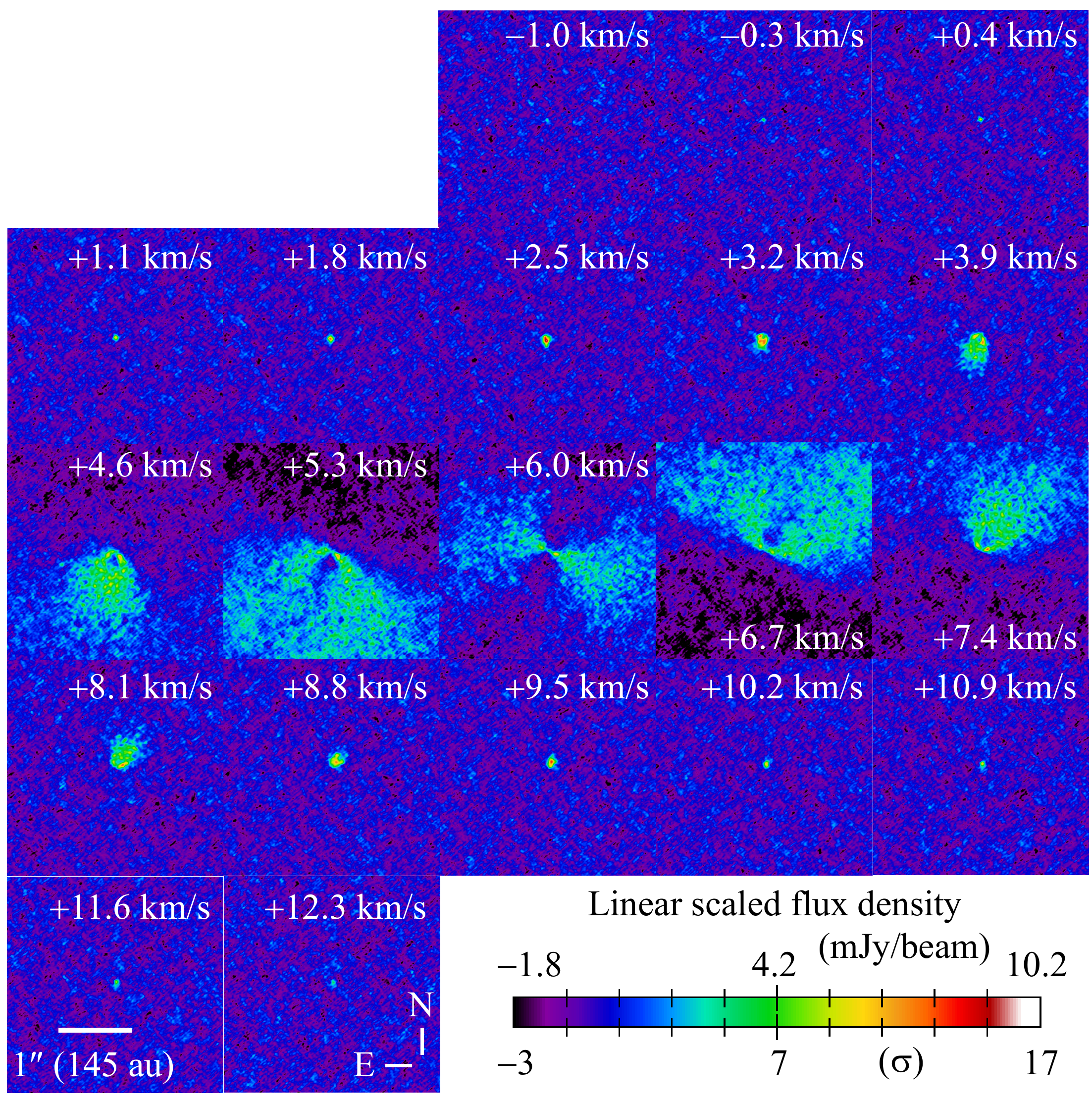}
\end{centering}
\caption{Channel maps for $^{12}$CO~$J=2\rightarrow1$. The r.m.s.\ noise for the 0.7-km/s bin is 0.589~mJy/beam with a beam size of 45.6~$\times$~45.4~mas and a PA of $-$77.9$^{\circ}$}\label{figA3}
\end{figure}

Figure~\ref{figA3} shows the $^{12}$CO~$J=2\rightarrow1$ channel maps at $-$1.0 to $+$12.3~km/s.

\section{Blob structure}

\begin{figure}[ht!]
\begin{centering}
\includegraphics[clip,width=\linewidth]{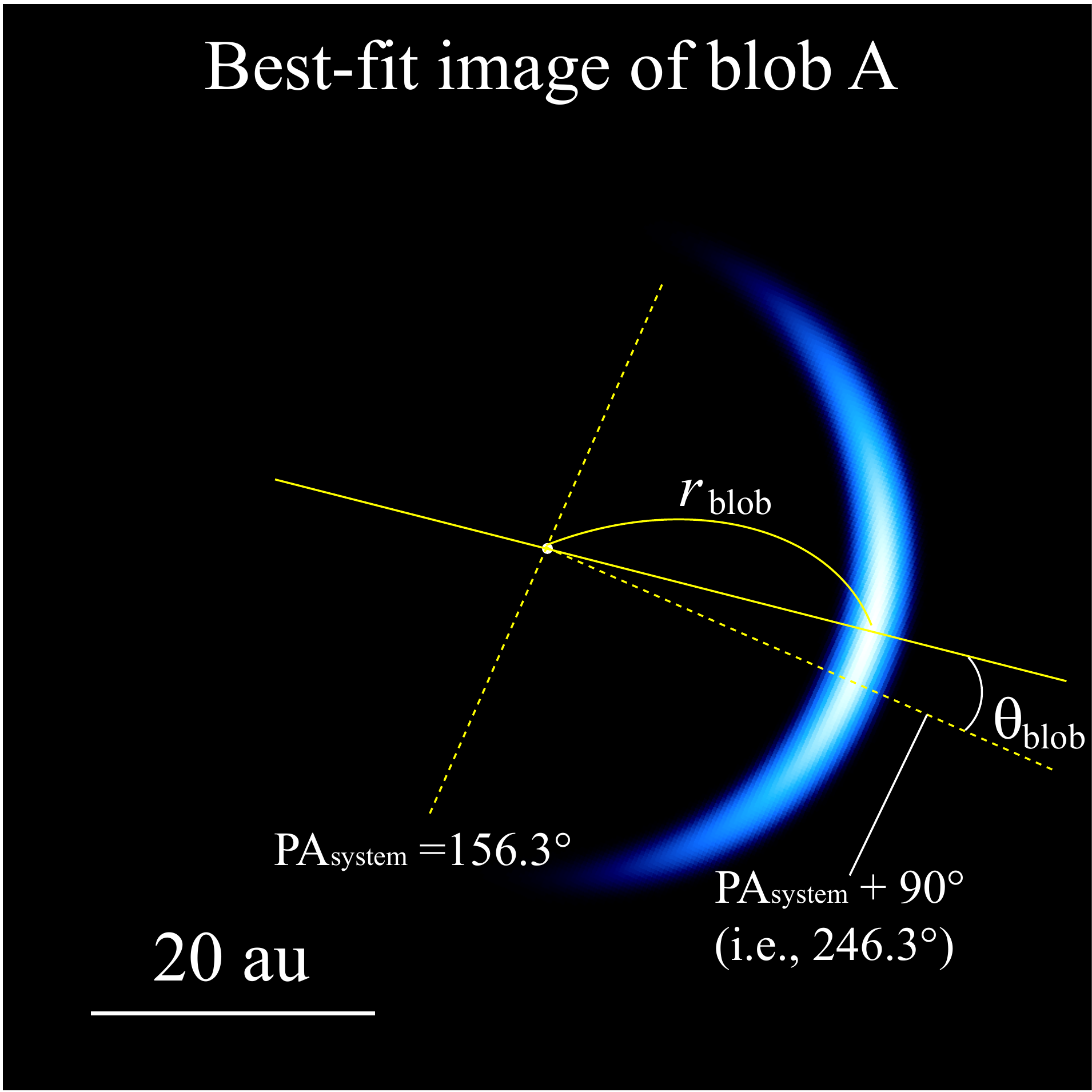}
\end{centering}
\caption{The best-fit model image of bolb~A derived by MCMC calculations in \S~\ref{subsec:mcmc}. Note that this image is not magnified with the system inclination of $i_{\rm system}=$36.1\degr.}\label{figA11}
\end{figure}

Figure~\ref{figA11} shows the model image of blob~A used in MCMC calculations in \S~\ref{subsec:mcmc}. 

\section{Weight values and the standard deviations in real and imaginary parts}

\begin{figure}[ht!]
\begin{centering}
\includegraphics[clip,width=\linewidth]{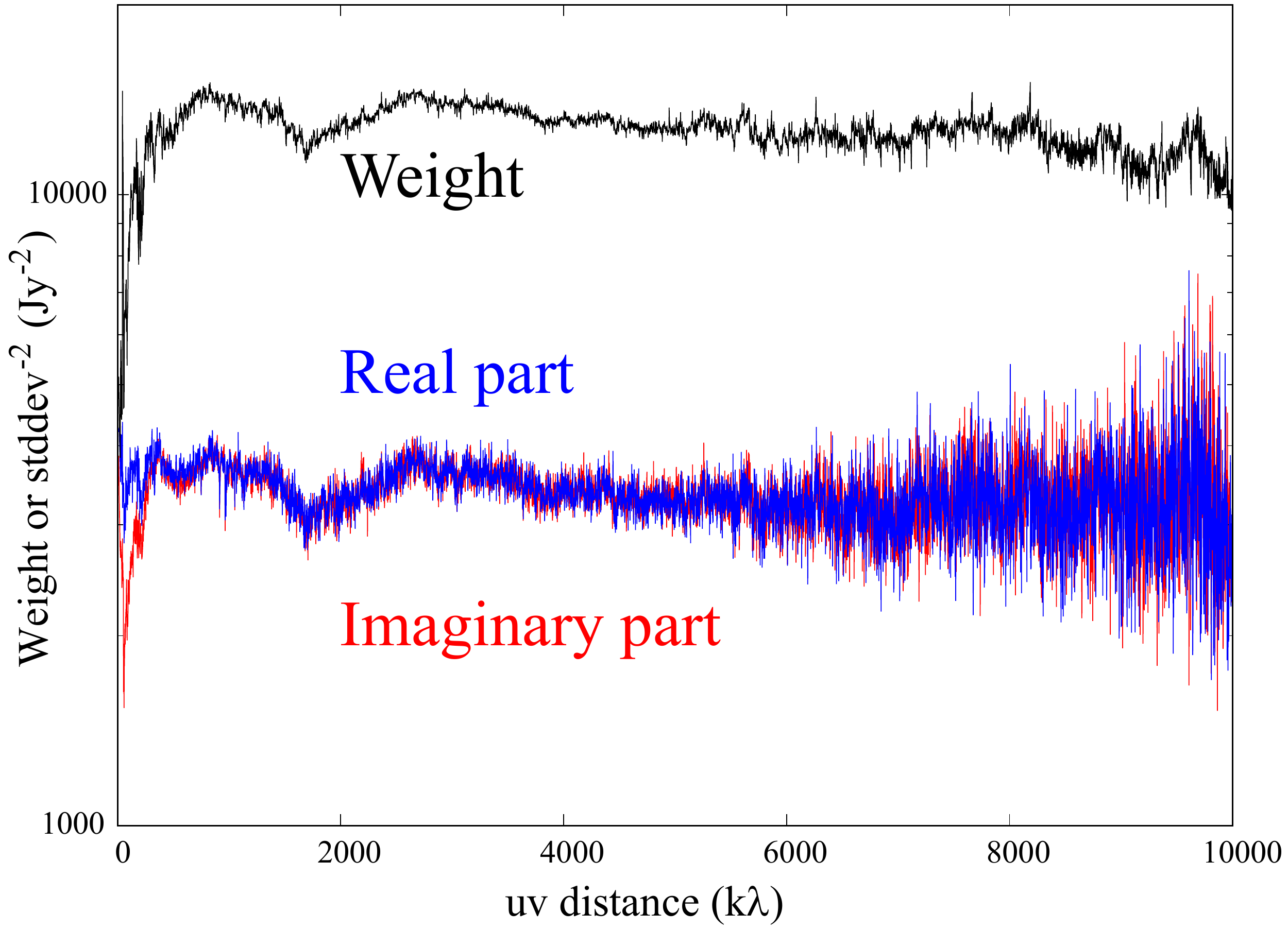}
\end{centering}
\caption{Comparison of the values of weight and the standard deviations (stddev) in real and imaginary parts. The values of weight are typically 3.85$\times$ higher than 1/stddev$^{2}$ of real and imaginary parts. 
}\label{figA8}
\end{figure}

Figure~\ref{figA8} shows the comparison of the values of weight and the standard deviations in real and imaginary parts. The standard deviations (stddev) are calculated in each 3~k$\lambda$~bin in the deprojected visibilities of real and imaginary parts with $i$ of 36\degr and PA of 156\degr. We found that the values of weight are typically 3.85$\times$ higher than 1/stddev$^{2}$ of real and imaginary parts. 

\section{Trace plot in MCMC calculations}

\begin{figure}[ht!]
\begin{centering}
\includegraphics[clip,width=\linewidth]{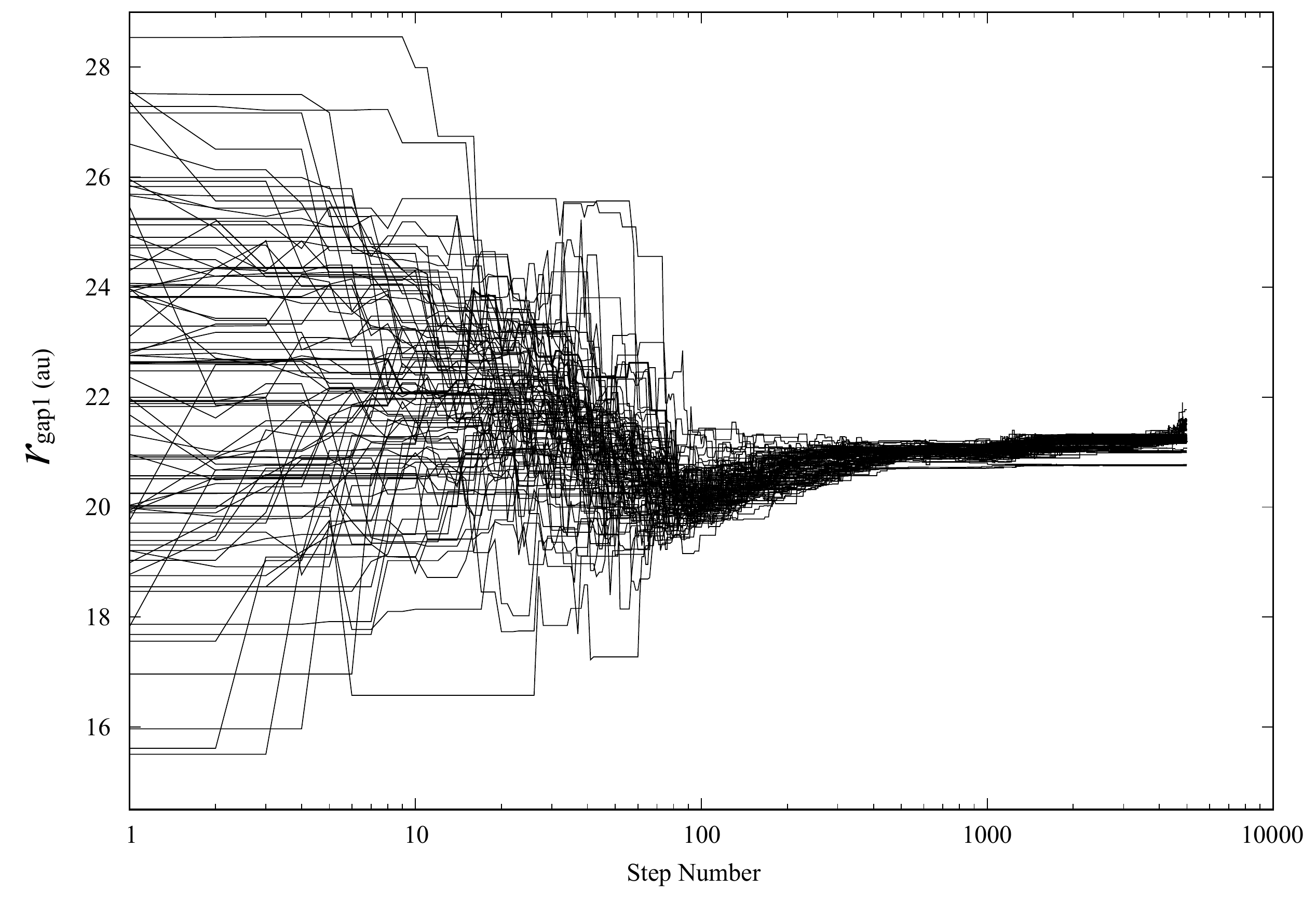}
\end{centering}
\caption{The trace plot of 100~walkers in the parameter $r_{\rm gap1}$. The initial 500~steps are set as the burnt-in phase and are discarded in the histograms of the marginal distributions of the MCMC posteriors in Figure~\ref{figA2}.
}\label{figA7}
\end{figure}

Figure~\ref{figA7} shows the trace plot of 100~walkers of the parameter $r_{\rm gap1}$ in our MCMC calculations (\S~\ref{subsec:mcmc}). The burnt-in phase is set as the initial 500~steps.

\section{Histograms of the marginal distributions of the MCMC posteriors}

\begin{figure}[ht!]
\center
\includegraphics[clip,width=0.85\linewidth]{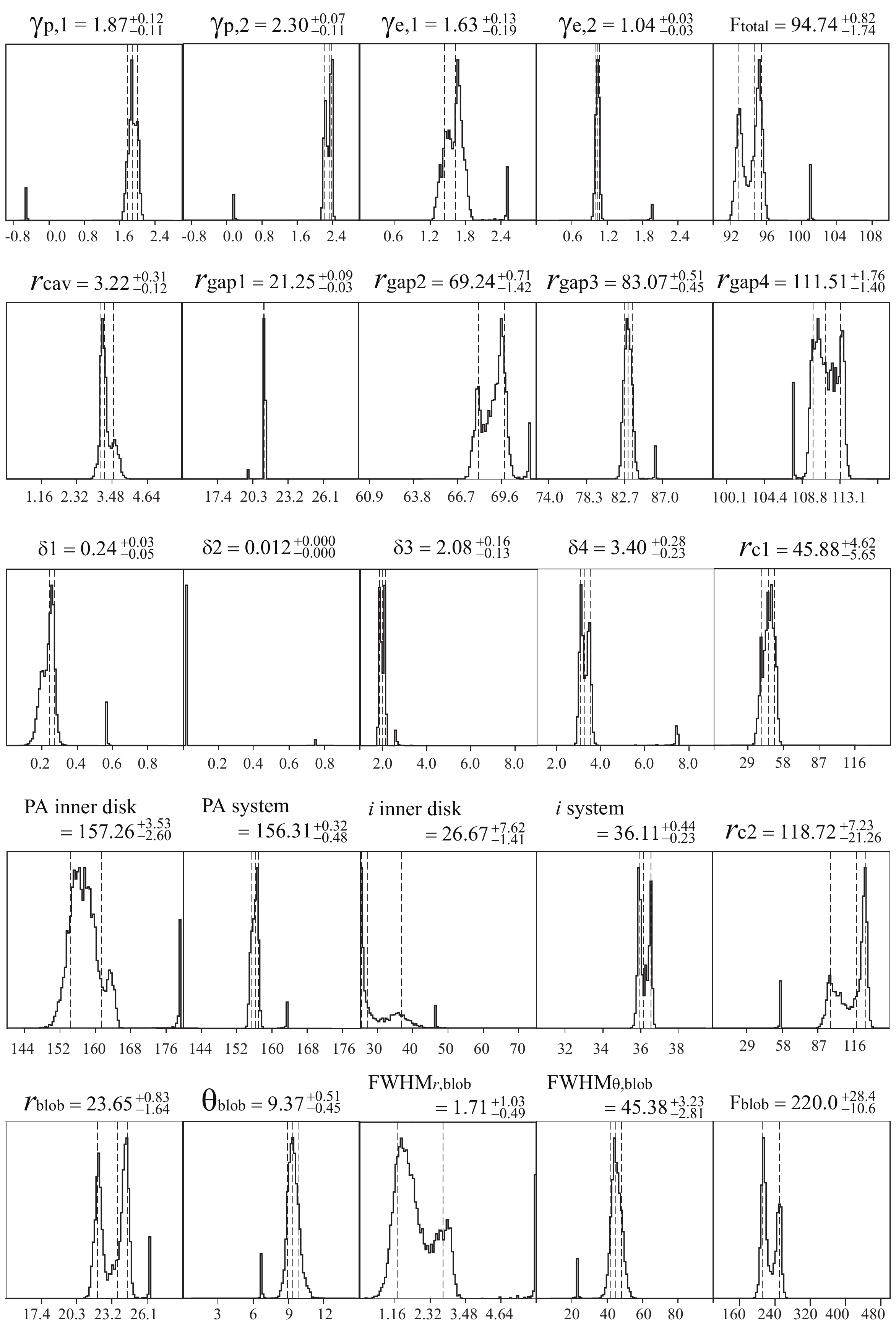}
\caption{
Histograms showing marginal distributions of the MCMC posteriors for 25 free parameters calculated in our modeling in \S~\ref{subsec:mcmc}. The best-fit values and their errors are computed from the 50th and the 16th and 84th percentiles, respectively, denoted as vertical dotted lines.
}\label{figA2}
\end{figure}

Figure~\ref{figA2} shows histograms of the marginal distributions of the MCMC posteriors for 25 free parameters calculated in our modeling in \S~\ref{subsec:mcmc}.

\section{A possible large scale asymmetry}

\begin{figure}[ht!]
\center
\includegraphics[clip,width=0.85\linewidth]{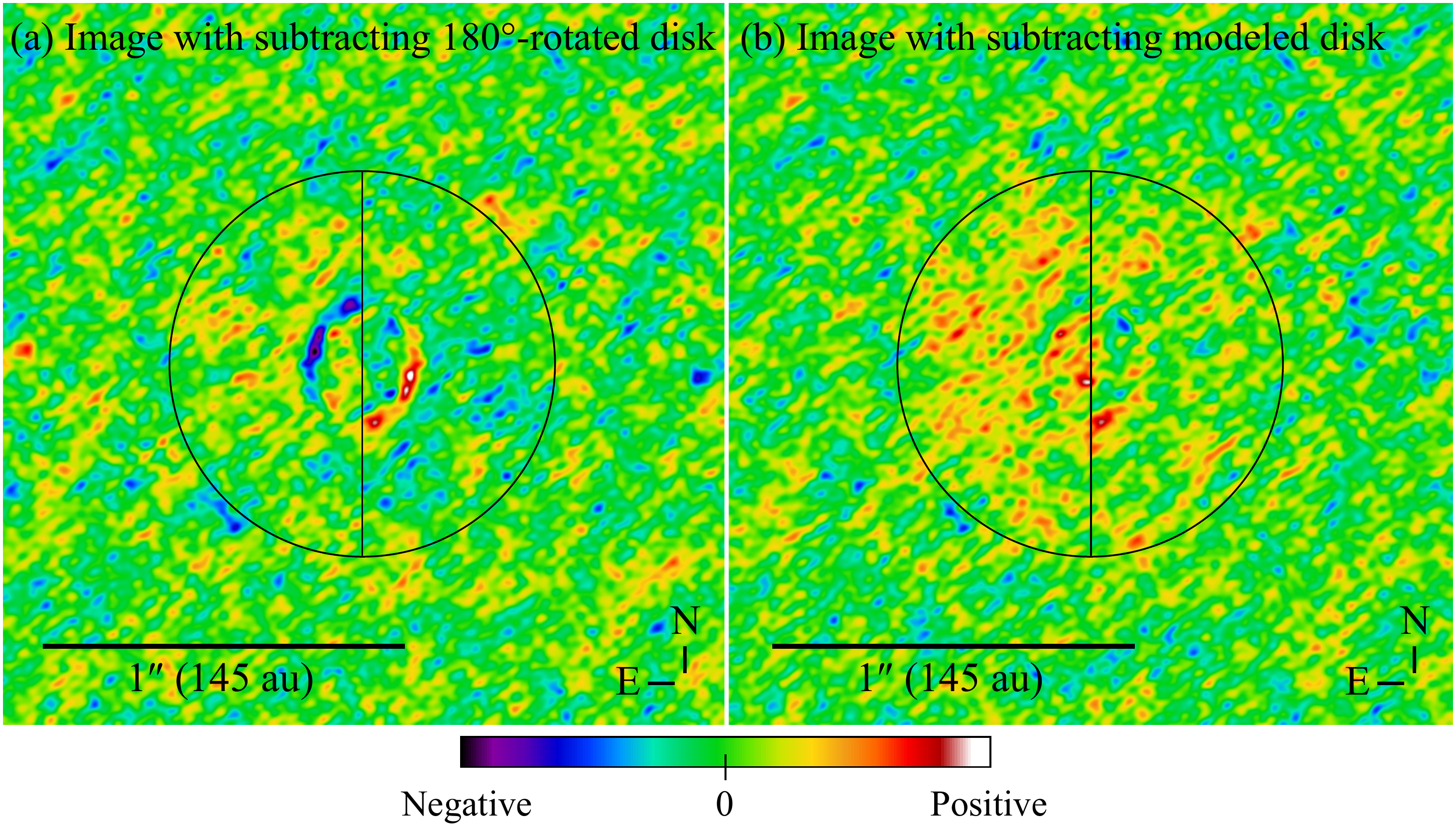}
\caption{
Two dust continuum images are taken from Figure~\ref{fig6}(a) and Figure~\ref{fig2}(e) to check a possible large scale asymmetry in the east part of the disk.
}\label{figA12}
\end{figure}

Figure~\ref{fig2}(e) in our visibility analyses in \S~\ref{subsec:mcmc} suggests that residual image subtracting the modeled disk is asymmetry in the east part. To check this asymmetry, the residual image is compared with the image subtracting the 180\degr-rotated image in Figure~\ref{figA12}. Both images show that the east part of the disk is brighter. Thus, DM~Tau could have the disk with large scale asymmetry. 

\section{Midplane temperature calculated by the MCRT modeling}

\begin{figure}[ht!]
\begin{centering}
\includegraphics[clip,width=\linewidth]{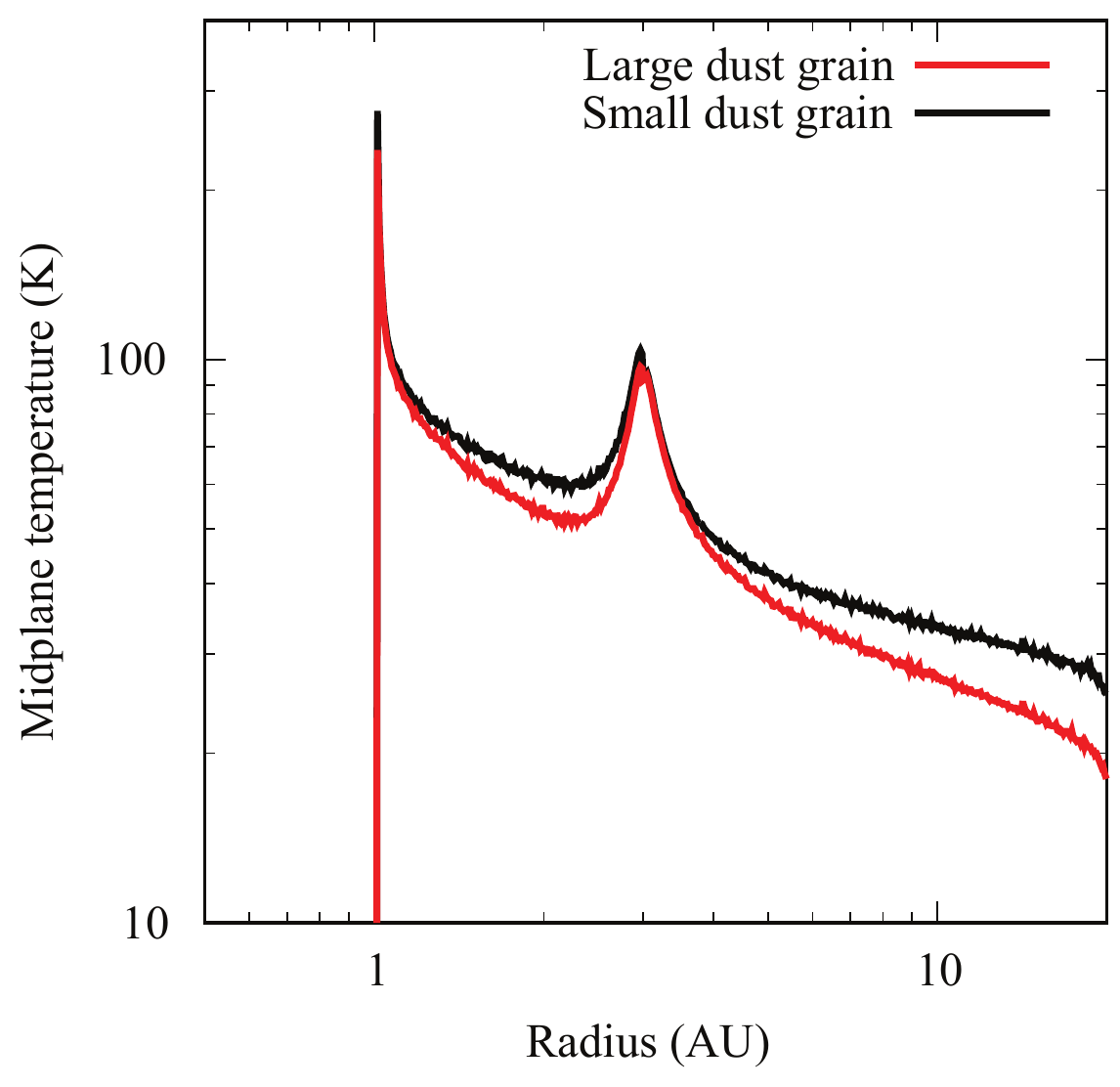}
\end{centering}
\caption{Midplane temperature of large and small dust grains in the fiducial model calculated in the MCRT modeling in \S~\ref{subsec:mcrt}}\label{figA6}
\end{figure}

Figure~\ref{figA6} shows a profile of the midplane temperature of large and small dust grains in the fiducial model calculated in the MCRT modeling in \S~\ref{subsec:mcrt}.

\section{The azimuthally averaged radial profiles generated by the MCMC model fitting and MCRT modeling}

\begin{figure}[ht!]
\begin{centering}
\includegraphics[clip,width=\linewidth]{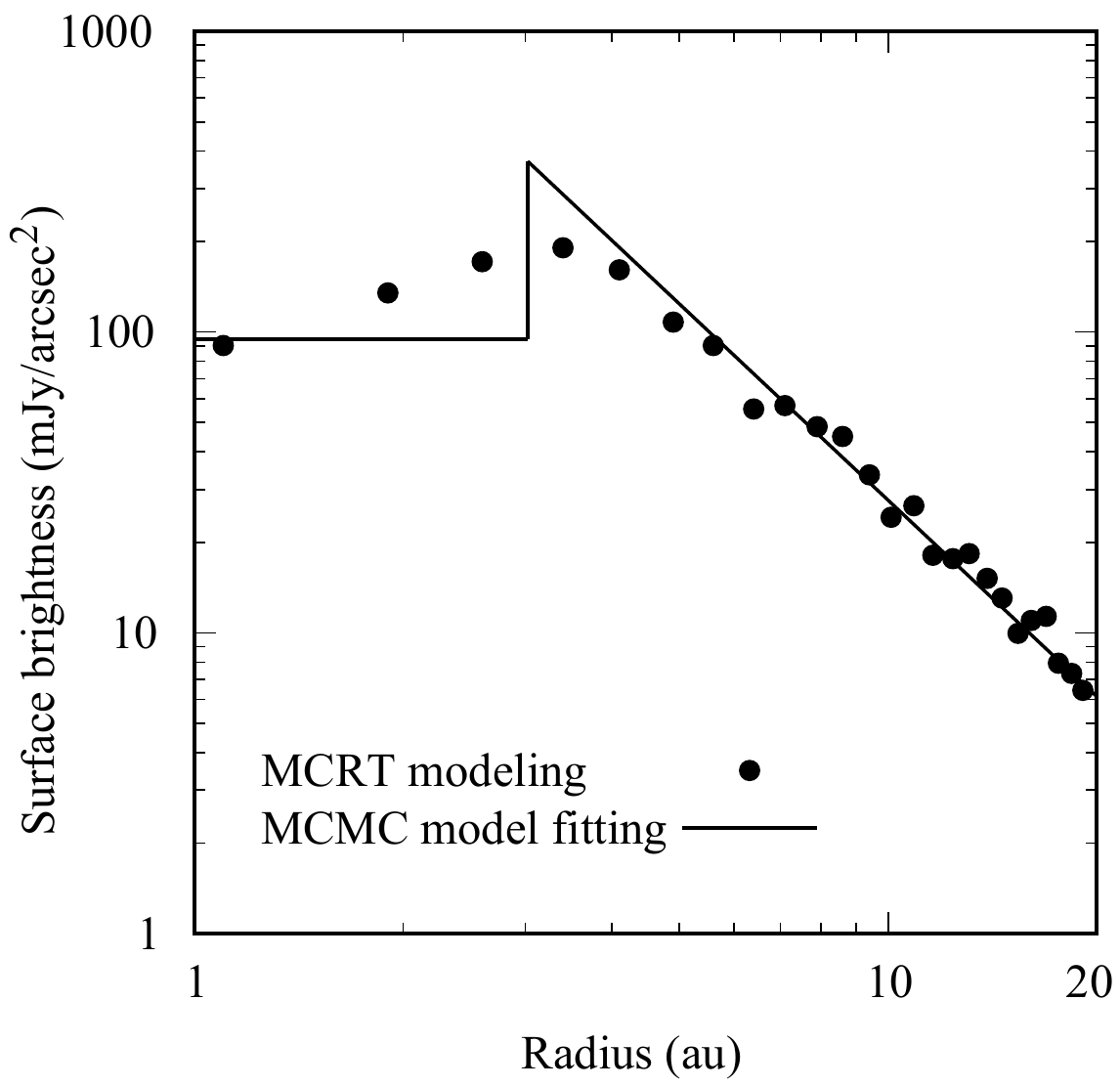}
\end{centering}
\caption{Radial profiles of modeled dust continuum images generated by MCMC model fitting and MCRT modeling. The profile in the MCMC model fitting at 3~au~$<r<$~20~au is described as 
$I(r) \propto \left(\frac{r}{45.88\ {\rm au}}\right)^{-1.87}{\rm exp}\left[-\left(\frac{r}{45.88\ {\rm au}}\right)^{1.63}\right]$ in \S~\ref{subsec:mcmc}.
}\label{figA4}
\end{figure}

Figure~\ref{figA4} shows the azimuthally averaged radial profile at $r<$ 20~au generated by the MCMC model fitting and MCRT modeling in \S~\ref{subsec:mcmc} and \S~\ref{subsec:mcrt}, respectively, to test the consistency of the surface brightness in the two modeling efforts.

\bibliography{arXiv20210211}{}
\bibliographystyle{aasjournal}

\end{document}